\documentclass[aps,prd,twocolumn,superscriptaddress]{revtex4-1} 
\usepackage{fullpage, amsmath, xparse, amssymb, mathtools, graphicx, braket,xcolor,verbatim}
\usepackage{natbib}
\usepackage[colorlinks=true,linkcolor=blue,citecolor=blue]{hyperref}

\newcommand{\mk}[1]{{\textcolor{green}{{\bf MK}: #1}}}
\newcommand{\aeber}[1]{{\textcolor{red}{{\bf AE}: #1}}}

\usepackage{multirow}
\usepackage{cancel}
\newcommand{\cev}[1]{\reflectbox{\ensuremath{\vec{\reflectbox{\ensuremath{#1}}}}}}

\begin{document}
\title{The classical field approximation of ultra light dark matter: quantum breaktimes, corrections, and decoherence}
\author{Andrew Eberhardt}
\thanks{Kavli IPMU Fellow. Corresponding author}
\email{\\ aeberhar@stanford.edu}
\affiliation{Kavli Institute for the Physics and Mathematics of the Universe (WPI), UTIAS, The University of Tokyo, Chiba 277-8583, Japan}
\author{Alvaro Zamora}
\thanks{Corresponding author}
\email{alvarozamora@stanford.edu}
\affiliation{Kavli Institute for Particle Astrophysics and Cosmology, Menlo Park, 94025, California, USA}
\affiliation{Physics Department, Stanford University, Stanford, California, USA}
\affiliation{SLAC National Accelerator Laboratory}
\author{Michael Kopp}
\affiliation{Nordita,
KTH Royal Institute of Technology and Stockholm University,
Hannes Alfv\'ens v\"ag 12, SE-106 91 Stockholm, Sweden}
\author{Tom Abel}
\affiliation{Kavli Institute for Particle Astrophysics and Cosmology, Menlo Park, 94025, California, USA}
\affiliation{Physics Department, Stanford University, Stanford, California, USA}
\affiliation{SLAC National Accelerator Laboratory}
\begin{abstract}

The classical field approximation is widely used to better understand the predictions of ultra-light dark matter. Here, we use the truncated Wigner approximation method to test the classical field approximation of ultra-light dark matter. This method approximates a quantum state as an ensemble of independently evolving realizations drawn from its Wigner function. The method is highly parallelizable and allows the direct simulation of quantum corrections and decoherence times in systems many times larger than have been previously studied in reference to ultra-light dark matter. Our study involves simulation of systems in 1, 2, and 3 spatial dimensions. We simulate three systems, the condensation of a Gaussian random field in three spatial dimensions, a stable collapsed object in three spatial dimensions, and the merging of two stable objects in two spatial dimensions. We study the quantum corrections to the classical field theory in each case. We find that quantum corrections grow exponentially during nonlinear growth with the timescale being approximately equal to the system dynamical time. In stable systems the corrections grow quadratically. We also find that the primary effect of quantum corrections is to reduce the amplitude of fluctuations on the deBroglie scale in the spatial density. Finally, we find that the timescale associated with decoherence due to gravitational coupling to Baryonic matter is at least as fast as the quantum corrections due to gravitational interactions. 
These results strongly imply that quantum corrections do not impact the predictions of the classical field theory.

\end{abstract}

\maketitle

\section{Introduction}

The standard model of cosmology, $\Lambda$CDM, is known to successfully describe much of the observed structure growth in the universe \cite{Planck2020}. This model includes a dark matter component comprising approximately $26\%$ of the universe's total energy budget. The observational evidence for dark matter is extensive, the distribution of mass in the bullet cluster \cite{Paraficz2016, Clowe2004}, the stellar rotation curves of galaxies \cite{sinclair1936, Zwicky1937}, and the anisotropies in the cosmic microwave background \cite{Planck2020, Spergel_2003} being  some of the most prominent examples. And while the density, self interaction, and temperature of the cold dark matter constituent of this model are constrained by observation, the specific particle nature remains unknown \cite{Feng2010}. This has motivated a large number of models spanning $\sim 100$ decades orders of magnitude of mass parameter space. 

At the lowest mass end, around $\lesssim 10^{-19} \, \mathrm{eV}$, we have ``ultra-light" dark matter (ULDM) models, see reviews \cite{Ferreira:2020fam,hui2021wave,SFDM_review}. Such ultra-light fields arise naturally in many string theory models \cite{Arvanitaki_2020}. Importantly, the low mass in this model means that the particles must be Bosonic \cite{Tremaine1979} and have a non-thermal production mechanism \cite{Viel_2013}. Here the mass of the particles is so low that their deBroglie wavelength is astrophysical in size \cite{Hu2000}. The deBroglie wavelength is given 
\begin{align}
    \lambda =  0.48\, \mathrm{kpc} \left( \frac{10^{-22} \, \mathrm{eV}}{m} \right) \left( \frac{250 \, \mathrm{km/s}}{v} \right) \, ,
\end{align}
in terms of the mass, $m$, and velocity, $v$, of the dark matter particle. Structures below the deBroglie scale are washed out while larger scale structures are left unchanged. It was originally hoped that a particle with mass $m\sim 10^{-22} \, \mathrm{eV}$ could alleviate small-scale structure problems without invoking Baryonic physics \cite{Hu2000}. These problems are usually summarized as the missing satellites \cite{Klypin:1999uc, Moore1999}, core-cusp \cite{navarro1996, Persic1995, Gentile2004}, and too-big-to-fail \cite{Boylan-Kolchin2011} problems, see \cite{Weinberg2015, Bullock2017} for review. Though the original $m\sim 10^{-22} \, \mathrm{eV}$ mass particle has been excluded by a large body of work, ultra-light dark matter remains an interesting model and work on this model helps establish a lower bound on the dark matter mass.

Ultra-light dark matter is associated with a rich phenomenology. Current constraints on ultra-light dark matter include the Lyman-$\alpha$ forest \cite{Armengaud2017, Irsic2017, Nori2018, Rogers2021}, galactic subhalo mass function~\cite{Nadler2021, Schutz2020}, stellar dispersion of ultra-faint dwarfs~\cite{Marsh:2018zyw,Dalal2022}, galactic density profiles~\cite{Bar2018,Bar2022, Zoutendijk2021}, Milky Way satellites~\cite{Safarzadeh2020}, gravitational lensing \cite{Powell2023}, and superradiance \cite{Arvanitaki2010, Davoudiasl:2019nlo}, see reviews \cite{Ferreira:2020fam,hui2021wave,SFDM_review} for specific details on constraint curves.

Typically, these constraints are derived by combining the predictions of theory and simulations to observations. Crucially, much of the analytic and numerical works relies on the classical field approximation, or mean field theory. In the full quantum field theory, quantum field operators representing the dark matter field act on a distribution of field values. In the classical field approximation, this distribution is replaced by the mean field value. This approximation is known to be accurate when the underlying quantum distribution is tightly peaked around the mean field value and when large occupation numbers mean the fractional variation in the field values due to ``quantum fluctuations", representing the non-zero width of the underlying distribution, are small. We generally say that the fractional deviation due to fluctuations goes as \cite{Guth2015}
\begin{align}
    \frac{\delta \hat \psi}{|\psi|} \sim \frac{1}{\sqrt{n_{tot}}} \, .
\end{align}
Where $\delta \hat \psi$ is a quantum field operator measuring the deviation from the mean field value, $\psi$, and $n_{tot}$ is the total number of particles in the system.

The masses considered are well below the thermal dark matter limit and so we need an alternative production mechanism for this model, for example the misalignment mechanism \cite{Preskill:1982, ABBOTT1983133}. The misalignment mechanism also has the advantage of placing the dark matter initially in a coherent state, making it amenable to classical approximation at early times. The light mass also means the occupation numbers are very large, a typical halo may have $n_{tot} \sim 10^{100}$ particles putting us well within the limit where quantum fluctuations are vanishingly small when the system is tightly distributed around the mean field value. This combination of 1) the initial coherent state description, and 2) the large occupation numbers of the system are typically used to justify the classical field description \cite{Guth2015, Hertberg2016}. 

In the absence of nonlinear interactions and when the initial conditions are well described classically, the accuracy of the mean field equations is known to survive, with canonical examples being Bose-Einstein condensates \cite{Leggett2001} and freely propagating photons \cite{Glauber1963}. However, it is known that nonlinearities, like those due to gravitational interactions, can introduce quantum corrections on some timescale, known as the ``quantum breaktime", even in highly occupied systems initially well described by classical theory \cite{Yurke1986, eberhardt2021Q, Eberhardt2022, Eberhardt_testing, TANAS1983351, Zurek2003, Zurek1998, Alon2007}. It has been shown that this is due to the underlying quantum distribution evolving away from one tightly peaked around the mean field value, for example \cite{Lentz2019, Lentz2020, Sikivie:2016enz, eberhardt2021Q, Eberhardt2022,Eberhardt_testing, Chakrabarty:2017fkd, chakrabarty2021, Hertberg2016, Yurke1986, TANAS1983351, Zurek2003, Zurek1998, Alon2007}. This deviation can be quantified by, or described as, a number of effects including the chaotic exploration of phase space \cite{Zurek2003}, phase diffusion \cite{Yurke1986}, fragmentation \cite{Alon2007}, etc. Importantly, it is not sufficient to simply compare the quantum breaktime to other timescales in the system. To fully understand the impact of quantum corrections it is also necessary to know the decoherence time and pointer states. The decoherence time is the timescale on which interactions with the environment 
entangle the quantum state with its basis of pointer states. Analytic estimates of the decoherence time indicate that it is short compared to the system dynamical times \cite{Allali2020, Allali2021, Allali2021b}. 

There has been a great deal of work examining the quantum nature of ultra-light dark matter as well as debate whether the classical field theory is sufficient to describe ultra-light dark matter on the scales relevant to constraints \cite{Sikivie2009, Guth2015, Dvali:2017eba, Dvali_2018, Allali2020, Lentz2019, Lentz2020, eberhardt2021Q, Eberhardt2022,Eberhardt_testing, Sikivie2012, kopp2021nonclassicality, chakrabarty2021, Chakrabarty:2017fkd, Hertberg2016, Marsh:2022gnf}, a more detailed description reviewing these works can be found in \cite{Eberhardt:2023sdb}. Some work has found that quantum corrections grow on the order of the dynamical time even in the highly occupied regime \cite{Sikivie:2016enz, Eberhardt2022, kopp2021nonclassicality, chakrabarty2021, Chakrabarty:2017fkd}. Others have argued that they remain small \cite{Guth2015, Allali2020, Allali2021, Allali2021b, Hertberg2016, Dvali_2018, Dvali:2017eba}. However, much of this work relies largely on analytic estimates, which may not be reliable into the nonlinear regime, or simulations of small toy systems, which may not be indicative of the behavior of more realistic ones.  

In previous works, we studied quantum corrections using full quantum simulations in small toy systems \cite{eberhardt2021Q}, and using the Field Moment Expansion method \cite{eberhardt2021} to study the gravitational collapse of an initial over-density in a single spatial dimension \cite{Eberhardt2022}. In both cases, we found that the gravitational interaction caused quantum corrections to grow exponentially. The latter case shows specifically that this growth occurred during the nonlinear growth of the over-density. Here we will use the truncated Wigner approximation \cite{Opanchuk2013, GonzlezArroyo2012, Sinatra2000, ruostekoski2012,Sinatra2002}. This method works by sampling the quantum Wigner distribution with many classical fields, i.e. an ensemble of solutions of the different realisations of the scalar field. A quantum state can be represented by its Wigner distribution \cite{Wigner1932,BIALYNICKIBIRULA2000,POLKOVNIKOV2010}. This can then be used to calculate the expectation value of operators corresponding to observables such as the field amplitude and spatial density. Its evolution can be approximated by an ensemble of classical fields all evolving according to MFT \cite{Hertberg2016, Opanchuk2013,POLKOVNIKOV2010}. By constructing this ensemble and simulating each constituent realization, an individual time-evolved classical representative of the ensemble,  in parallel we can accurately and quickly simulate the evolution of observables. Already this method has been successfully used to simulate Bose-Einstein condensates in a trap \cite{Opanchuk2013,Sinatra2000,ruostekoski2012,Sinatra2002}, quantum number eigenstates with a four point interaction \cite{Hertberg2016}, gravitational collapse of an initial overdensity in a single spatial dimension for a coherent state \cite{Eberhardt_testing}, and quantum field theory calculations \cite{GonzlezArroyo2012} among others.

We study the ``quantum breaktime" of ultra light dark matter using the method presented in \cite{Eberhardt2022}, i.e. measuring the growth rate of the $Q$ parameter, a proxy for the spread of the wavefunction around its mean field value. We find that our three dimensional results corroborate the one dimensional results in \cite{Eberhardt2022}. The main result being that quantum corrections grow exponentially during nonlinear collapse and halo merging, but only quadratically after merging. We also study the impact of large quantum corrections on the evolution of the density. We find that large quantum corrections tend to remove the $\sim \mathcal O(1)$ density fluctuations at the deBroglie scale . Finally, we study decoherence using a test particle intended to represent Baryonic matter which we know to take well defined trajectories through phase space. We find that these test particles quickly enter into  macroscopic superpositions, spreading around their mean value at the same rate as the dark matter wavefunction. This strongly implies that the macroscopic super positions needed to impact the predictions of the the classical field theory do not occur in realistic systems. The results of this paper support the conclusion that the classical field theory produces accurate predictions for scalar field dark matter. However, a more complete answer would require the identification of pointer states which is beyond the scope of this paper.

This work in organized in the following way: in Section \ref{sec:backgroundMSM} we discuss the necessary quantum mechanical background. In Section \ref{sec:TWA_MSM} we discuss the truncated Wigner approximation. Section \ref{sec:Test_problems_MSM} describes the test problems we simulate. We summarize results in Section \ref{sec:results_MSM} and discuss their implications in Section \ref{sec:discussion_MSM}. Finally, we conclude in Section \ref{sec:conclusions_MSM}.

\section{Background} \label{sec:backgroundMSM}

In this section we introduce the quantum field and quantum phase space formalisms. We then show how the classical field theory is derived in the limit of large occupation number and when the quantum distribution is tightly peaked around the mean field value. Following this, we describe how quantum corrections enter the system when these assumptions are relaxed. Finally, we introduce the decoherence formalism.  

\subsection{Quantum description}

In the non-relativistic limit, the Hamiltonian of a self gravitating quantum scalar field takes the following form

\begin{align} \label{eqn_Ham}
    \hat H / \tilde \hbar &= \sum_j \omega_j \hat a_j^\dagger \hat a_j + \sum_{ijkl} \frac{\Lambda_{kl}^{ij}}{2} \hat a_k^\dagger \hat a_l^\dagger \hat a_i \hat a_j  \\
    &= \iint dx dy \, \hat \psi^\dagger(x) \frac{-\tilde \hbar \,  \nabla^2}{2} \hat \psi(y) +  \hat \psi^\dagger(x) \, \frac{\hat V(x)}{\tilde \hbar} \, \hat \psi(x) \, ,  \nonumber
\end{align}
where $\tilde \hbar = \hbar / m$. In our case the potential will be the solution to Poisson's equation, $\nabla^2 \hat V(x) = C m \, \hat \psi^\dagger(x) \hat \psi(x)$. The position and momentum space field operators, $\hat \psi(x)$ and $\hat a_i$ respectively, are related by Fourier transform

\begin{equation} \label{psi2a}
    \hat \psi(x) = \sum_i \hat a_i u^\dagger_i(x) \, . 
\end{equation}

$\hat a$ and $\hat a^\dagger$ are the annihilation and creation operators respectively. $u^\dagger_i(x)$ is the momentum eigenstate with momentum $\hbar k_i$. The field operators act on a quantum state, $\ket{\psi}$. We will be concerned with the time evolution of $\ket{\psi}$. It is often convenient to write this state in the basis of number eigenstates $\set{\ket{n_i}}$ which satisfy $\hat a^\dagger \hat a \ket{n} = n \ket{n}$. Where then $\ket{n_i}$ is the number eigenstate with $n$ particles in the $i$th momentum mode. 

We can analysis this system by looking at the evolution of the quantum state, given by Schr\"odinger's equation
\begin{align}
    i \hbar \, \partial_t \ket{\psi} = \hat H \ket{\psi} \, ,
\end{align}
or by looking at the evolution of the quantum field operators, given by Heisenberg's equation
\begin{align}
    i \hbar \, \partial_t \hat \psi = [\hat \psi, \, \hat H] \, .
\end{align}

In this work we focus mainly on the simulation of coherent states which we can write the following 

\begin{equation} \label{eqn_coherentStates}
    \ket{\Vec{z}}_C = \bigotimes_{i=1}^M \exp \left[ -\frac{|z_i|^2}{2} \right] \sum_{n_i=0}^\infty \frac{ z_i^{n_i}}{\sqrt{n_i!}} \ket{n_i} \, .
\end{equation}

where $\vec z$ is the vector of Fourier components of the classical field, i.e. $z(x) = \sum_i z_i u_i^\dagger(x)$. This type of state is thought to describe the initial conditions for ultra-light dark matter produced via the misalignment mechanism \cite{ABBOTT1983133, Preskill:1982}. 

\subsection{Phase space representation and pseudo probability distributions}

In much of this work, it will be much more convenient to work in phase space. The representation of operators and states in phase space is described by their Weyl symbols. For an arbitrary operator,$\hat \Omega(\set{\hat \psi, \hat \psi^\dagger})$, which is a function of our set of field operators $\set{\hat \psi, \hat \psi^\dagger}$, the Weyl symbol is given

\begin{widetext}
\begin{align}
    \Omega_W[\psi, \psi^*] \equiv \frac{1}{\mathrm{Norm}} \int_{\mathbb{C}^{\mathbb{R}^3}} \int_{\mathbb{C}^{\mathbb{R}^3}} \mathcal{D} \eta \, \mathcal{D} \eta^* \braket{ \psi - \frac{\eta}{2} \, | \, \hat \Omega(\set{\hat \psi, \hat \psi^\dagger}) \, | \, \psi + \frac{\eta}{2} }_C e^{-|\psi|^2 - \frac{1}{4}|\eta|^2} e^{\frac{1}{2}(\eta^* \psi - \eta \psi^* )} \, .
\end{align}
where $\mathcal{D} \eta = \Pi_x d \eta(x)$
denotes a functional integral measure over all complex field configurations, see \cite{OpanchukDrummond2013} for a rigorous treatment. 
For operators which are symmetrically ordered functions of $\hat \psi, \hat \psi^\dagger$ the Weyl symbol can be found by making the substitution $\hat \psi, \hat \psi^\dagger \rightarrow \psi, \psi^*$ in $ \hat \Omega(\set{\hat \psi, \hat \psi^\dagger})$. The Weyl symbol of this operator is a real valued functional of the field configuration $\psi(x), \, \psi^*(x)$.

We will make use of the Wigner function, $f_W$, which is the Weyl symbol of the density matrix, $\hat \rho$. The Wigner function defines a pseudo probability distribution on this phase space. It is not a true probability distribution because it takes negative values for most states. 

The Weyl symbol of the commutator, $[\dots \, , \, \dots]$, is the Moyal bracket which acts as

\begin{align}
    \set{\!\set{f,g}\!}_M \equiv 2 f(\psi, \psi^*) \sinh \left( \frac{1}{2} ( \cev{\partial}_\psi \vec{\partial}_{\psi^*} - \cev{\partial}_{\psi^*} \vec{\partial}_{\psi} ) \right) g(\psi, \psi^*) \, .
\end{align}
Note that when the amplitude of $\psi$ is large compared to the higher order derivatives of the Wigner function, i.e. $|\partial_\psi^3 f|/|\partial_\psi f| \ll n_{tot}$, 
that the Moyal bracket can be approximated as a Poisson bracket, i.e.  

\begin{align}
    \set{\!\set{f,g}\!}_M = \set{f,g}_c + \mathcal{O}(1/n_{tot}) \, .
\end{align}

We can then write the Von Neumann equations of motion as 

\begin{align} \label{eqn:phaseSpaceSchr}
    \partial_t \hat \rho &= -\frac{i}{\hbar} \left[ \hat H, \hat \rho \right] \rightarrow \\
    \partial_t f_W[\psi, \psi^*;t] &= -\frac{i}{\hbar} \left\{ \left\{ H_W[\psi, \psi^*]\, , \, f_W [\psi, \psi^*;t] \right\} \right\}_M \, \\
    &\approx -\frac{i}{\hbar} \left\{ H_W[\psi, \psi^*]\, , \, f_W [\psi, \psi^*;t]  \right\}_c \, . \label{eqn:phaseSpaceShcrApprox} 
\end{align}

Expectation values are calculated 

\begin{align}
        \braket{\hat \Omega(\set{\hat \psi, \hat \psi^\dagger})} &= 
\int_{\mathbb{C}^{\mathbb{R}^3}} \int_{\mathbb{C}^{\mathbb{R}^3}} \mathcal{D} \psi \, \mathcal{D} \psi^* \,  f_W[\psi, \psi^*] \, \Omega_W[\psi, \psi^*] \, . 
\end{align}
\end{widetext}

For our purposes one particuarly important Wigner function is that of a coherent state, see equation \eqref{eqn_coherentStates}. This distribution is simply Gaussian centered on the mean field value \cite{OpanchukDrummond2013}, i.e. for a coherent state with classical field $z(x)$
\begin{align} \label{eqn:wignerCoherent}
    f_W[\psi, \psi^*] = \frac{1}{\pi} e^{-\int dx |\psi(x) - z(x) |^2} \, .
\end{align}

\subsection{Classical field approximation}

Using the phase space and pseudo probability distribution formalism it is straightforward to understand the assumptions necessary to justify the classical field approximation. We first identify the classical field as the mean field value, $\psi^{cl}(x) = \braket{\hat \psi(x)}$, with an initial state given by equation \eqref{eqn:wignerCoherent} we have $\psi^{cl}(x,t=0) = z(x)$.. Then we solve Heisenberg's equation in the phase space for $\partial_t \braket{\hat \psi(x,t)}$. 

\begin{widetext}

\begin{align}
    \partial_t \braket{\hat \psi(x,t)} &= \partial_t \psi^{cl}(x) = -\frac{i}{\hbar} \int_{\mathbb{C}^{\mathbb{R}^3}} D\psi \, \left\{ \set{H_W[\psi, \psi^*]\,,\,\psi(x)} \right\}_M\, f_W[\psi, \psi^*] \, .
\end{align}

Next we take the large occupation limit $n_{tot} \gg 1$, which allows us to approximate the Moyal bracket as a Poisson bracket,

\begin{align}
    \partial_t \psi^{cl}(x) \approx -\frac{i}{\hbar} \int_{ \mathbb{C}^{\mathbb{R}^3} } D\psi \, \left\{ H_W[\psi, \psi^*]\,,\,\psi(x) \right\}_c\, f_W[\psi, \psi^*] \, .
\end{align}

Next we assert that the distribution is tightly peaked around the mean field value, i.e. $|\braket{\hat \psi}|^2 \gg \braket{\delta \hat \psi^\dagger \, \delta \hat \psi}$; the distribution can therefore be approximated as a delta function at the classical field value, $f_W = \delta[\psi(x) - \psi^{cl}(x)]$

\begin{align}
    \partial_t \psi^{cl}(x) &\approx -\frac{i}{\hbar} \int_{\mathbb{C}^{\mathbb{R}^3}} D\psi \, \left\{ H_W[\psi, \psi^*]\,,\,\psi(x) \right\}_c\, \delta[\psi - \psi^{cl}] \, \\
    &= -\frac{i}{\hbar} \left\{ H_W[\psi^{cl}, \psi^{cl*}]\,,\,\psi^{cl}(x) \right\}_c \, , \\
    &= -\frac{i}{\hbar} \left( \frac{\nabla^2}{2m} + m\,V \right) \psi^{cl}(x)
\end{align}

\end{widetext}

The last line is the familiar Schr\"odinger-Poisson equations when $\nabla^2 V(x) = C m | \psi^{cl}(x)|^2$. We see that the derivation of the classical field equations of motion relies on two assumptions. The first is that the Moyal bracket was well approximated by a Poisson bracket, which is true in the large occupation limit, i.e. $n_{tot} \gg 1$. The second is that the quantum distribution is tightly localized around the classical field value, i.e. $|\braket{\hat \psi}|^2 \gg \braket{\delta \hat \psi^\dagger \, \delta \hat \psi}$, see discussions in \cite{eberhardt2021Q, eberhardt2021, Sikivie:2016enz}. Because these assumptions are necessary for the classical field equations but will be relaxed in the following sections, it will be useful later to parameterize the size of quantum corrections due to the spreading of the wavefunction using $Q(t)$, see \cite{eberhardt2021}.

\begin{align} \label{eqn:Q}
    Q(t) = \frac{1}{n_{tot}} \int dx \, \braket{\delta \hat \psi^\dagger(x) \delta \hat \psi(x)} 
\end{align}

Note that a coherent state with large classical field amplitude, see equation \eqref{eqn:wignerCoherent}, satisfies both of the necessary assumptions for description by a classical field and has $Q = 0$. If the state remains well approximated by a coherent state with large occupations then the classical field theory will remain accurate \cite{eberhardt2021Q, eberhardt2021, Eberhardt2022}. 

\subsection{Quantum corrections}

\begin{figure*}[!ht]
	\includegraphics[width = .97\textwidth]{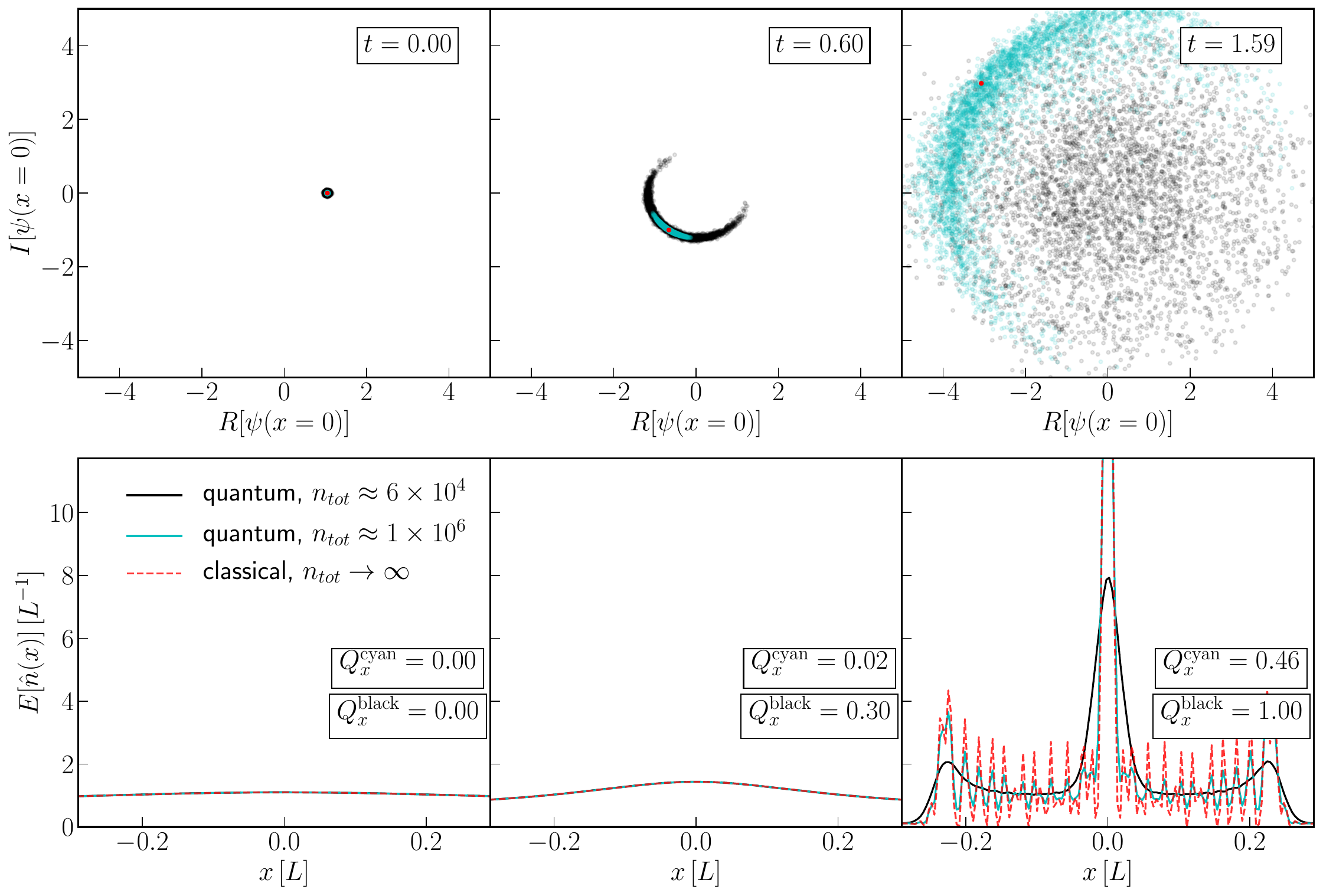}
	\caption{ Gravitational collapse of a spatial over-density in a single spatial dimension. We plot the results for the classical field theory in red, and two quantum simulations of coherent states with $n_{tot} \approx 6 \times 10^4$ and $n_{tot} \approx 1 \times 10^6$ in black and cyan, respectively. Each column represents a different time, $t$. The top row shows the value of the each stream in the ensemble at $x=0$ and the bottom row shows the spatial density plotted such that each field has the same norm. Shell crossing occurs at $t=1$. During the collapse phase the field undergoes phase diffusion but the density remains well approximated by the MFT until after the collapse. Following the collapse, the density is smoothed out in proportion to the amount of phase diffusion achieved prior to the collapse, with high particle number simulations exhibiting larger fluctuations in the final number density. In these simulations $\tilde \hbar = 2.5 \times 10^{-4}$ and $N_s = 1024$, $M=256$, $M_{tot} = L = 1$. Plot taken from \cite{Eberhardt_testing}. }
	\label{fig:1D_density_compare}
\end{figure*}

Quantum corrections begin to enter the system when the assumptions used to derive the mean field theory to break down. As discussed in the previous section, the classical field theory is achieved in the limit where the Moyal bracket can be replaced by the Poisson bracket and the quantum distribution can be approximated by a delta function centered at the mean field value. The first approximation relies on the how the mean field values compare with higher order derivatives of the Wigner function. Corrections to the Poisson bracket approximation are of $\sim \mathcal{O}(1/n_{tot})$ \cite{POLKOVNIKOV2010} and are not discussed in this work. The second approximation ignores the finite width of underlying quantum distribution. This width is given by the commutation relation between the field operators $\hat \psi^\dagger$, $\hat \psi$, and can be related to the uncertainty principle and is therefore a correction $\sim \mathcal{O}(1/\sqrt{n_{tot}})$. This is the main correction considered in this work. 

Coherent states with large occupation numbers are thought to describe ultra light dark matter produced by the misalignment mechanism at early times \cite{ABBOTT1983133, Preskill:1982}. In the top row of Figure \ref{fig:1D_density_compare}, we plot the underlying quantum distribution as compared with the classical field value of the gravitational collapse of an initial over-density, for two quantum simulations of coherent states at different occupation number, $n_{tot}$, but with the same mean field evolution. We see in the top left panel of Figure \ref{fig:1D_density_compare}, that at early times the distribution is tightly peaked around the mean field value. Overtime, if the Hamiltonian has nonlinearities, like gravity, the underlying quantum distribution will spread. This is caused by the finite width of the distribution which creates correction terms to the equations of motion proportional to the variance of the field operators, see \cite{eberhardt2021,Eberhardt2022}. In the top middle panel of Figure \ref{fig:1D_density_compare}, the distribution is beginning to experience phase diffusion \cite{Yurke1986}. This means that the phase of wavefunction is accumulating uncertainty and is becoming less well defined. At shell crossing, this phase diffusion becomes amplitude uncertainty, that is, the distribution spreads around the ring in the complex plane of fixed amplitude corresponding to $A(x) = \sqrt{n_{tot}(x)}$, see the top right panel of Figure \ref{fig:1D_density_compare}. The position of $\sim \mathcal{O}(1)$ density fluctuations require well defined phase gradients. The result is that the density admits quantum corrections at late times, although well described by mean field theory at early times.

We parameterize these deviations by $Q(t)$, defined in equation \eqref{eqn:Q}, which give an approximate description of the leading order quantum corrections to the mean field equation. We will then define a ``quantum breaktime" as $Q(t_{br}) \sim 1$. This is the time when quantum corrections are large, and the system can no longer be well described by the classical field theory alone. 

It has been demonstrated that quantum corrections grow exponentially for chaotic systems, and, as a result, breaktimes scale logarithmically with occupation number \cite{Berman1978,Berry1979, Zurek1998, Hertberg2016, eberhardt2021Q, Kovtun2022, Eberhardt2022, Zurek2003}. The phase space description of quantum mechanics offers an intuitive explanation for this phenomenon. The distribution over quantum phase space can be thought, to first order, as a classical ensemble of fields with slight perturbations in their initial conditions. The chaotic quality of the system then causes these perturbations to exponentially grow apart in phase space, causing the distribution to spread away from its mean field value. The relationship between chaos and quantum phase space is explored in detail in \cite{Zurek2003}.

\subsection{Decoherence}

Let us consider the state, $\ket{\psi(t}$, of a system we are interested in, e.g. the dark matter halo of a galaxy. We couple this state to an environment, $\ket{\mathcal{E}(t)}$, e.g. the state describing the phase space of the stars in the galaxy. We will assume at our initial time, $t=0$, that the state describing both system and environment can be written as a product

\begin{align} \label{eqn:ICs}
    \ket{A(0)} &= \ket{\psi(0)}\ket{\mathcal{E}(0)} \, , \\ 
    &= \sum_i c_i(t=0) \ket{\phi}_i \otimes \sum_j b_j(t=0) \ket{\epsilon}_j \, .
\end{align}

The combined state evolves via the Hamiltonian

\begin{align}
    \hat{H}_A = \hat H_\psi + \hat H_\mathcal{E} + \hat H_{\mathrm{int}} \, .
\end{align}

Where $\hat H_{\mathrm{int}}$ describes the interaction between the state and environment. Assuming that the Hamiltonian is time independent, the evolution of the state to an arbitrary time $t = T$, is given 

\begin{align}
    \ket{A(T)} = e^{-i \, \hat H_A \, T} \ket{A(0)} \, .
\end{align}

At this time, because of the influence of the interaction term in our Hamiltonian, there is no guarantee that the state can be written as a simple tensor product as in equation \eqref{eqn:ICs}. In general, the system will be entangled with the environment. We now must write our state more generally as

\begin{align}
    \ket{A} = \sum_{ij} c_{ij} \ket{\phi}_i \ket{\epsilon}_j \, . 
\end{align}

Which can describe the entanglement between the two sets of basis states. It will often be convenient beyond this point to write the state's density matrix

\begin{align}
    \hat \rho_A = \sum_{ijkl} c_{ij} c^*_{kl} \ket{\phi}_i \ket{\epsilon}_j \bra{\phi}_k \bra{\epsilon}_l \, .
\end{align}

Now, if we assert that an observer measures the environment to be in the eigenbasis $\ket{\tilde \epsilon}$. We then have a reduced density matrix tracing over the environment eigenbasis

\begin{align}
    \hat \rho^R_A = \mathrm{Tr}_\epsilon[\rho_A] = \sum_i \bra{\tilde \epsilon} \hat \rho_A \ket{\tilde \epsilon}_i \, .
\end{align}

When $\braket{\tilde \epsilon| \tilde \epsilon}_{ij} = \delta_{ij}$, the reduced density matrix is now a classical ensemble of pointer states of the system. Pointer states being the states which develop the least entanglement over time with the preferred environmental basis states. This process of environmental interaction projecting the state into the pointer state basis is called ``decoherence". It is necessary to calculate the timescale associated with this process in order to fully understand the impact of quantum corrections. 

\section{Truncated Wigner approximation} \label{sec:TWA_MSM}

In this section we introduce the truncated Wigner approximation (TWA). We explain how we implement this scheme and how it can be used to model the quantum breaktime and decoherence. 

\subsection{Approximation scheme}

The truncated Wigner approximation scheme is a method for approximating the evolution of the Wigner function, equation \eqref{eqn:phaseSpaceSchr}, in a way that relaxes the assumption that the quantum distribution is tightly distributed around the classical field value. We can use this to estimate the leading order quantum corrections to the classical field theory.  

The TWA method relies on two sets of approximations. First, we approximate the time evolution in phase space dropping all terms of order $\mathcal{O}(1/n_{tot})$ and higher, i.e. 

\begin{align}
    \set{\!\set{f,g}\!}_M \approx \set{f,g}_c \, .
\end{align}

Note that this is the same first assumption required in the derivation of the classical field theory. The next approximation is of the Wigner function itself. We will represent the Wigner function with a set of classical fields and weights $\set{c_i, \psi_i(x)}_W$ as 

\begin{widetext}

\begin{align} 
    f_W[\psi, \psi^*;t] &\approx f_S[\psi, \psi^*;t] \\ 
    f_S[\psi, \psi^*;t] &= \frac{1}{N_s} \sum_i c_i \, \delta[\psi - \psi_i(x,t)] \, \delta[\psi^* - \psi_i^*(x,t)] \label{eqn_stream_ensemble}
\end{align}

Where $\psi_i(x,t)$ represents the $i$th field configuration in the set with weight $c_i$ and $N_s$ is the total number of streams in the set. We choose the field instances at $t=0$ such that

\begin{align}
    \int_{ C \subset \mathbb{C}^{\mathbb{R}^3}} D\psi \, f_W[\psi, \psi^*;t = 0] &={\lim_{N_s \rightarrow \infty} }\int_{C \subset \mathbb{C}^{\mathbb{R}^3}}  D \psi \, f_S[\psi, \psi^*,t = 0] \nonumber\\
    &= \lim_{N_s \rightarrow \infty} \frac{1}{N_s} \sum_i \begin{cases}
c_i &\psi_i(x, t= 0) \in C,\\
0 &\text{else} \, ,
\end{cases} 
\end{align}
for all regions $C \subset \mathbb{C}^{\mathbb{R}^3}$. Note that the above scheme is sufficiently general to include Wigner function which are not everywhere positive. If, however, the Wigner function being approximated is everywhere positive, it is sufficient to treat it as a probability distribution functional for the fields, i.e.  $\psi_i \sim f_W(\psi, \psi^*)$ with $c_i = 1$ for all $i$.

The expectation value of a symmetrically ordered operator at time $t$, $\braket{\hat \Omega[\set{ \hat\psi, \hat \psi^\dagger}]}$ is then given 

\begin{align} \label{eqn:E[op]}
    \braket{\hat \Omega(\set{ \hat\psi, \hat \psi^\dagger})} &= \int_{C \subset \mathbb{C}^{\mathbb{R}^3}} D\psi \, f_W[\psi, \psi^*,t] \, \Omega_W[\psi, \psi^*] \\
    &= \lim_{N_s \rightarrow \infty} \frac{1}{N_s} \sum_i c_i \, \Omega_W[\psi = \psi_i(x); t ] \, .
\end{align}

The accuracy of the truncation of the Moyal bracket as a Poisson bracket relies on large occupation numbers, $n_{tot} \gg 1$, like the classical field approximation. However, unlike the classical field approximation, we relax the assumption that the underlying quantum distribution is well approximated by a delta function at the mean field value. The TWA instead requires that we have adequate field instances sampled to resolve the distribution.

The equation of motion for the set is given as 
\begin{align}
    \partial_t f_S[\psi, \psi^*; t] &= -\frac{i}{\hbar N_s} \sum_i c_i  \set{H_W[\psi_i(x,t), \psi^*_i(x,t)]\, , \, \psi_i(x,t)}_c \, , \\ 
    &= -\frac{i}{\hbar N_s} \sum_i c_i \left( \frac{\hbar^2\nabla^2}{2m} + m\,V_i(x,t) \right) \psi_i(x,t)
\end{align}
\end{widetext}
Achieved by plugging equation \eqref{eqn_stream_ensemble} into equation \eqref{eqn:phaseSpaceShcrApprox}. It is important to note the somewhat unintuitive result that, at this approximation order, the potential, $V_i$, is a functional only of $\psi_i$ not the ensemble of fields, i.e. $\nabla^2 V_i(x) = C m | \psi_i(x)|^2$. The equation of motion for each individual stream is given 

\begin{align}
    \partial \psi_i(x,t) = -\frac{i}{\hbar} \left( \frac{\hbar^2\nabla^2}{2m} + m\,V_i(x,t) \right) \psi_i(x,t)
\end{align}

This result is derived in more detail in appendix \ref{apndx:proof}. 
Notice also, that each stream evolves independently of any other, meaning that this method is highly parrallelizable. 


\subsection{Numerical implementation}

\begin{figure*}[!ht]
	\includegraphics[width = .97\textwidth]{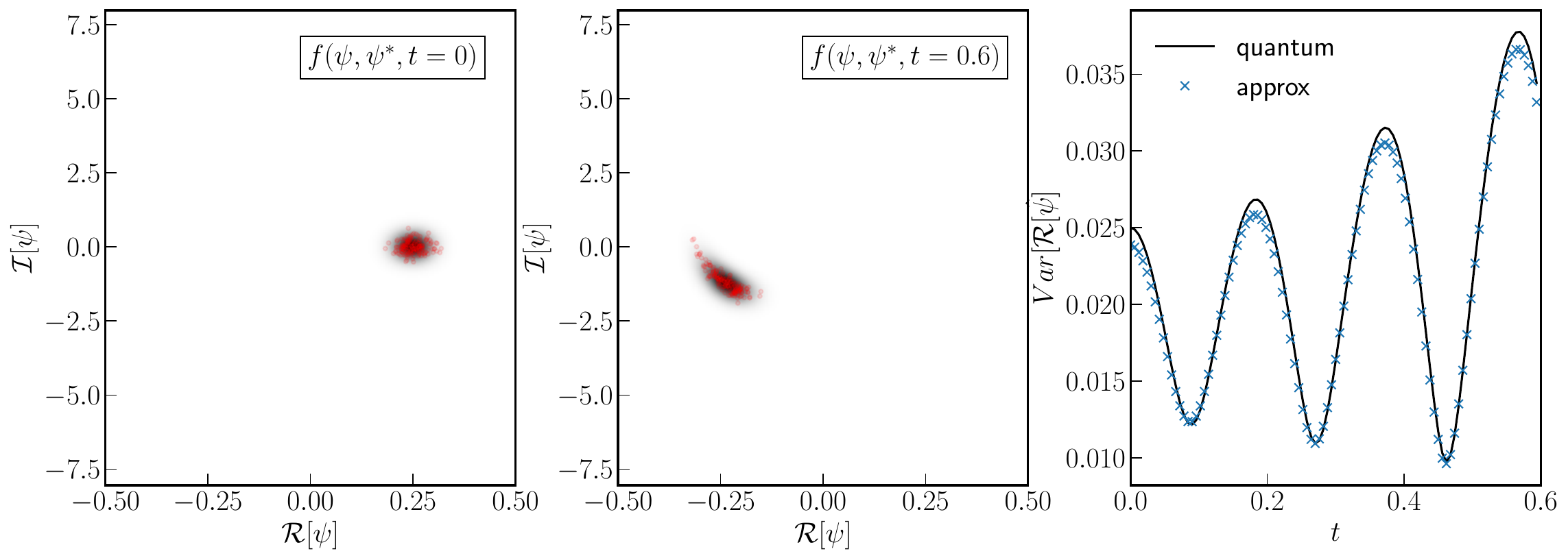}
	\caption{ Here we show how the approximation scheme works diagrammatically for a single mode with a quartic nonlinearity. The left two plots show the Husimi distribution for the quantum distribution at the initial and final times. The red dots show the value of the many random streams sampling the Wigner function of the quantum distribution. We can see that the evolution of the ensemble of points approximates the evolution of the underlying quantum phase space. The right most plot shows the true value and ensembled approximation of $\mathrm{Var}[\hat q]$. We can see that the evolution is well approximated by the ensemble. Here we use $N_s = 1024$. }
	\label{fig:MSM_diagram}
\end{figure*}

Numerically, we solve this system by integrating the mean field evolution of an ensemble of classical fields instances. This means that any solver which solves the Sch\"odinger-Poisson equations can be used. Because each of the streams evolves independently of the others, we can solve each in parallel, allowing for a large number of streams to be simulated efficiently. We use the pseudo spectral integrator, as well as timestep and resolution checks described in \cite{Eberhardt2020} with an updated kinetic aliasing check \footnote{\label{footnote:alias} Contrary to the typical kinetic time step constraint $dt < \frac{8 \pi}{\tilde \hbar k_{max}^2}$ typically used in the literature, we use $dt < \frac{4 \pi}{\tilde \hbar k_{max} \, dk}$, where $k_{max} = \hbar / 2 / dx$ is the maximum momentum mode in the system and $dk$, and $dx$ are the momentum and position space resolutions respectively. That is, we prevent the gradient of the phase from aliasing rather than the phase itself at any given point in the field. This restriction is a more precise one since only the phase gradients are meaningful. This provides a large speed up in our simulations over the conventional time step. }. A discussion of the language used for the implementation as well as the code repository link are in appendix \ref{appendixB}.

\subsubsection{Initial conditions generation}

We will simulate the evolution of coherent states, see equation \eqref{eqn_coherentStates}. The Wigner function for a coherent state is a Gaussian centered on the classical field value, $\psi^{cl}(x)$, see equation \eqref{eqn:wignerCoherent}, i.e. the exact Wigner function at the initial conditions for a coherent state is 
\begin{align}
    f_W[\psi, \psi^*] = \frac{1}{\pi} e^{-\int dx |\psi(x) - \psi^{cl}(x)|^2} \, .
\end{align}
We will approximate this with a stream ensemble as
\begin{align}
    f_S[\psi, \psi^*] &= \frac{1}{N_s} \sum_i c_i \, \delta[\psi - \psi_i(x)] \, \delta[\psi^* - \psi^*_i(x)] \, .
\end{align}

Where each field instance is drawn from the true Wigner function, $\psi_i(x) \sim f_W[\psi, \psi^*]$. Note that for a coherent state, the Wigner function has an untroubled interpretation as a probability distribution. Here it is also necessary to introduce our spatial grid, which in three spatial dimensions is written $x_{ijk} = (i \, dx, \, j \, dx, \, k \, dx)$ where $dx = L/M$ is the spatial resolution given by the box size, $L$, over the number of spatial modes in a single dimension $M$. For notational convenience, we write the grid indices in a way independent of the number of spatial dimensions, e.g. let $ijk = g$ where now $g \in \set{0,1,\dots, M-1}^D$, where $D$ is the number of spatial dimensions. Then we choose our fields as

\begin{align}
    \psi_i(x_g) = \psi^{cl}(x_g) + \delta^R_i(x_g) + i \, \delta^I_i(x_g)  \, .
\end{align}

Where at every point, $g$, in our grid we choose two random numbers drawn from a Gaussian distribution with variance $1/2$, i.e.  $\delta^R_i(x_g), \, \delta^I_i(x_g) \sim \mathcal{N}(0, \sqrt{1/2})$. Note that this is only the case if the classical field is normalized to be the number density, i.e. $\sum_g|\psi(x_g)|^2\, dx = n_{tot}$. 
Let us define a normalized $\psi'(x) \equiv \psi(x) / \sqrt{n_{tot}}$ such that $\sum_g|\psi'(x_g)|^2\, dx = 1$, as is often more convinient, then
\begin{align} \label{eqn:sample}
&\psi'_i(x_g) = \psi'^{cl}(x_g) + \delta'^R_i(x_g) + \delta'^I_i(x_g) i \, , \\
&\delta'^R_i(x_g), \, \delta'^I_i(x_g) \sim \mathcal{N}(0, \sqrt{1/2}) / \sqrt{n_{tot}}. 
\end{align}
See \cite{Opanchuk2013} for a more detailed discussion of this sampling scheme.

\subsubsection{Integrating the equations of motion}

The fields are integrated using the standard symplectic pseudo spectral leap frog integrator following the temporal and spectral aliasing resolutions constraints discussed in \cite{Eberhardt2020,Garny}, however we update the kinetic temporal resolution check.

Let $\psi_t^i \equiv \psi_i(x,t)$, $V_t \equiv V(x,t)$, and $\tilde \psi \equiv \mathcal{F}[ \psi]$, i.e. the Fourier transform of the field. The update $\psi_t \rightarrow \psi_{t+\Delta t}$ is given in the non-expanding case

\begin{align*}
    \Tilde{\psi}_{t+\Delta t} &= U^T_t(\Delta t/2)\Tilde{\psi}_t \textrm{  (position update half step)} \\
    &\textrm{(calculate $V_t$)} \\ 
    \psi_{t+\Delta t} &= U^V_t(\Delta t)\psi_{t} \textrm{  (momentum update full step)} \\
    \Tilde{\psi}_{t+\Delta t} &= U^T_t(\Delta t/2)\Tilde{\psi}_t \textrm{  (position update half step)} \, .
\end{align*}

$U^T$ and $U^V$ are the unitary operators associated with the kinetic and potential energies respectively, i.e. 

\begin{align}
    U^T_t(\Delta t) &\equiv e^{i \,\Delta t \hbar \, k^2 / (2m)} \, , \\
    U^V_t(\Delta t) &\equiv e^{- i \,\Delta t  \, m \, V(x,t)  / \hbar} \, .
\end{align}

$\Delta t$ is dynamically chosen to avoid temporal aliasing in the kinetic or potential updates. This means that at each time step 

\begin{align}
    \Delta t = 2 \pi c_f \, \mathrm{min} \left[ \hbar / m V \, , \, m L / (M \pi \hbar) \right] \, . 
\end{align}

Where the first argument of the minimum function is the restriction on the time step set by the potential energy and the second is the restriction set by the gradient of the kinetic energy. Notice that this differs from the restrictions in \cite{Eberhardt2020}, here we only ensure that the gradient of the kinetic energy does not alias.

\subsubsection{Evaluating operators}

The expectation value of symmetrically ordered operators, i.e. equation \eqref{eqn:E[op]}, can be approximated using our ensemble of fields as 

\begin{align}
    \braket{\hat \Omega[\set{ \hat \psi, \hat \psi^\dagger}]} \approx \frac{1}{N_s} \sum_{i} \, \Omega_W[\psi_i, \psi^*_i;t] \, .
\end{align}
Where now the elements of our set of operators are the field operators defined at the grid points, i.e. $\hat \psi(x_g) \in \set{\hat \psi, \hat \psi^\dagger}$ for all $x_g$.

\subsection{Estimating the breaktime} \label{subsec:breaktime}

\begin{figure*}[!ht]
	\includegraphics[width = .97\textwidth]{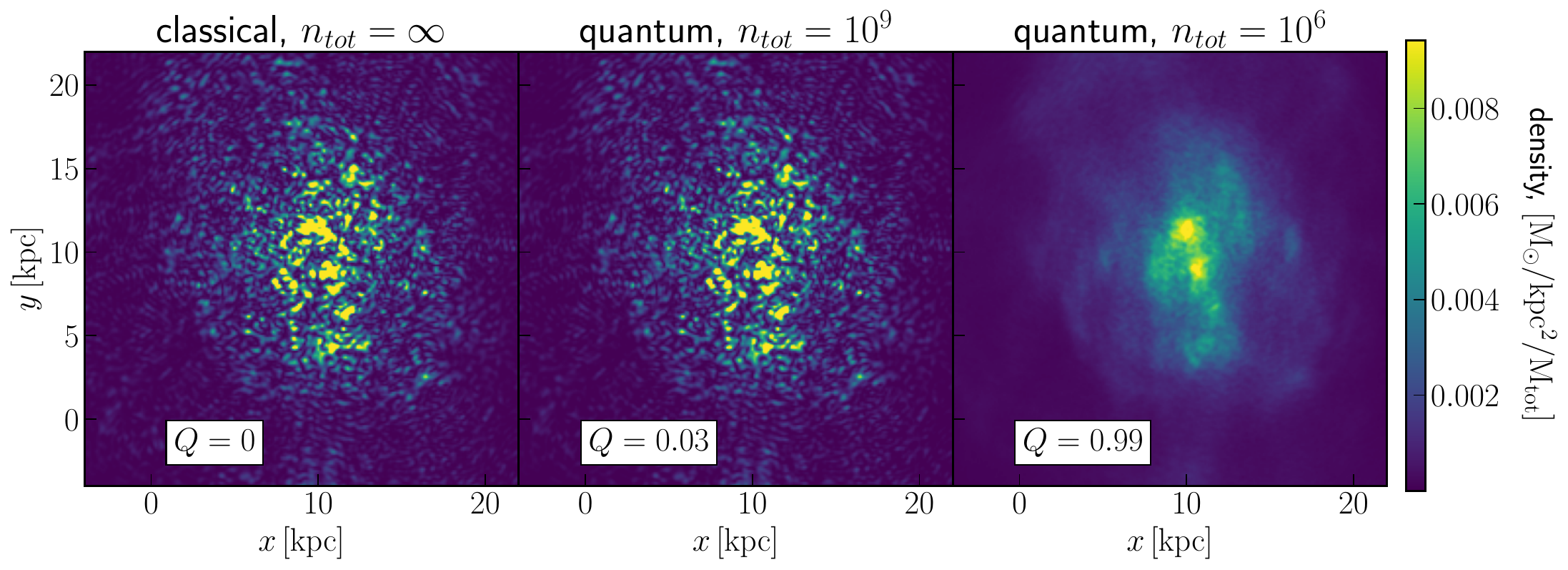}
	\caption{ Here we plot the condensed object resulting from the collapse of a momentum space Gaussian density in two spatial dimensions, initial conditions described in Section \ref{subsec:boseStar}. The density shown here is the result of $1 \, \mathrm{Gyr}$ of evolution. Each plot also shows the simulated value of the $Q$ parameter at this time. The left panel shows the classical field evolution, the condensed object in this case exhibiting the expected granular structure. The right panel shows the evolution for a simulation using the truncated Wigner approximation with $n_{tot} = 10^6$ particles. Here, as in the one dimensional case, we see that the quantum corrections have removed most of the granular structure. In the middle panel we show the same simulation with $n_{tot} = 10^9$ particles. The quantum corrections in this case are much smaller, and the resulting density is almost identical to the classical case. Here we set $M_{tot} = 6 \times 10^9 \, M_\odot$, $\hbar / m = 0.02 \, \mathrm{kpc^2/Myr}$, $L = 60 \, \mathrm{kpc}$, $M = 512^2$, $2 k_d^2 = 0.05 \, \mathrm{kpc^{-2}}$, $T = 1 \, \mathrm{Gyr}$. }
	\label{fig:2d_collapseQuantum}
\end{figure*}

There are a number of physical parameters in our simulations which determine the evolution of the system. For example, the total mass in the system $M_{tot}$, the mass of the field $m$, Planck's constant $\hbar$, and the total number of particles $n_{tot}$. In real systems, these parameters are related in physical ways, for example $m = M_{tot} / n_{tot}$. For the cosmological systems we are interested in, these parameters have large values. A typical halo may have $M_{tot} \sim 10^{10} \, M_\odot$, $m \sim 10^{-22} \, \mathrm{eV/c^2}$, $n_{tot} \sim 10^{100}$. If we use these values, the sampling scheme described in this section will fail as we do not have the numerical precision to model a $\delta \psi \sim 1/\sqrt{n_{tot}} \sim 10^{-50}$. We therefore simulate systems with non-physical values of these parameters and extrapolate them to physical values. We will describe this procedure in this section. Note that this is a similar procedure as previous work \cite{Eberhardt2022}.

We instead define the simulation parameters using their relation to the classical field evolution and the sampling scheme. If we normalize the fields $\psi$ such $\sum_g|\psi'(x_g)|^2\, dx = 1$ then we can write the classical Schr\"odinger-Poisson field equations as 

\begin{align} \label{eqn:mft_normed}
    \partial_t \psi'(x) &= -i \left( -\frac{\tilde \hbar \nabla^2}{2} + \frac{V(x)}{\tilde \hbar} \right) \psi'(x) \, , \\ 
    \nabla^2 V(x) &= 4 \pi G M_{tot} |\psi'(x)|^2 \, .
\end{align}

The classical field equations then depend only on our choice of $M_{tot}$ and $\tilde \hbar$ which give the total mass in the simulation and the mass of the field respectively. Notice that $n_{tot}$ does not enter the classical field equations, as expected since the classical equations are in the $n_{tot} \rightarrow \infty$ limit. Now $n_{tot}$ only enters as a sampling parameter in equation \eqref{eqn:sample} and is not defined by $M_{tot}/m$. From here, we distinguish between the physical value of $n^p_{tot} \equiv M_{tot}/m$,  and the simulated value of $n^s_{tot}$ which enters only in the field sampling $\delta'^R_i(x_g), \, \delta'^I_i(x_g) \sim \mathcal{N}(0, \sqrt{1/2}) / \sqrt{n^s_{tot}}$. It is interesting to note that a classical field simulation has $n^s_{tot} \ne n^p_{tot}$ and instead sets $n^s_{tot} = \infty$.

The parameter $Q(t)$, defined in \eqref{eqn:Q}, is a measure of quantumness \cite{eberhardt2021,eberhardt2021Q,Eberhardt2022,Eberhardt_testing}. When $Q(t) \ll 1$ then the system is well described by the classical field theory and when $Q(t) \sim 1$ then quantum corrections will begin to cause deviations from the classical theory, see for example Figures \ref{fig:2d_collapseQuantum} and \ref{fig:1D_density_compare}, which demonstrate the relationship between $Q$ and the density predcited by the quantum and classical evolutions. In this paper we will define a quantum breaktime, $t_{br}$, to be $Q(t_{br})\sim 1$. 

\begin{figure}[!ht]
	\includegraphics[width = .44\textwidth]{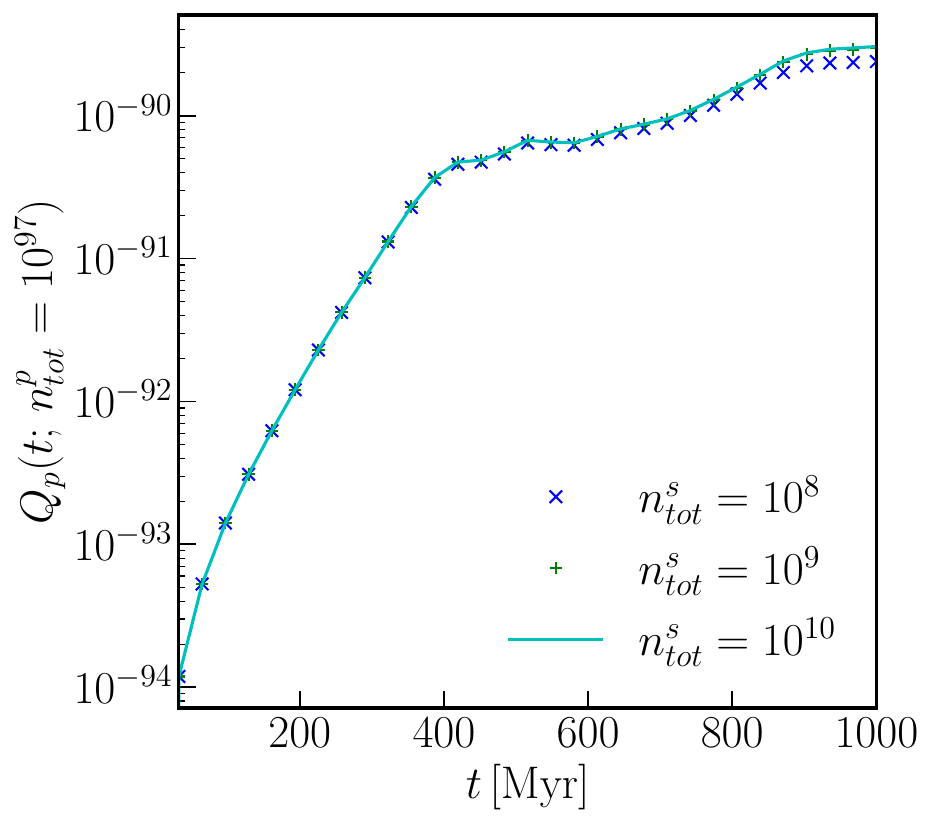}
	\caption{ Here plot the prediction for the physical $Q_p$ for three different simulations all with different sampling parameters $n^s_{tot}$. The simulated $Q_s(t)$ are then normalized according to equation \eqref{eqn:Q_relation}. We can see that each simulation makes the same prediction for the growth of $Q_p$. The simulations are of the collapse of a momentum space Gaussian, described in Section \ref{subsec:boseStar}, in two spatial dimensions, with $M_{tot} = 6 \times 10^9 \, M_\odot$, $\hbar / m = 0.02 \, \mathrm{kpc^2/Myr}$, $L = 60 \, \mathrm{kpc}$, $M = 512^2$, $2 k_d^2 = 0.05 \, \mathrm{kpc^{-2}}$. }
	\label{fig:Q_n_compare}
\end{figure}

However, as we just discussed, we do not have access to the true value of $Q_p(t)$ for a given set of physical parameters $M_{tot}$, $\hbar$, $m$, $n_{tot}$ that respect the relationship $n_{tot} = M_{tot} / m$. Instead, we have a simulated value of $Q_s(t; M_{tot}, n_{tot}, \tilde \hbar)$ where $M_{tot}$ and $\tilde \hbar$ specify the classical field evolution and $n_{tot}$ is only a sampling parameter. For fixed $M_{tot}$, $\tilde \hbar$, $Q$ with different $n_{tot}$ are related by the ratio of their respective $n_{tot}$ when $Q(t) \ll 1$. We can describe this relation as 

\begin{align} \label{eqn:Q_relation}
    Q_1(t; M_{tot}, \tilde \hbar, n_{tot} = n_1) = \frac{n_2}{n_1} Q_2(t; M_{tot}, \tilde \hbar, n_2) \, .
\end{align}

An example is instructive. Let us say that we want to model the evolution of the physical system with $M_{tot} = 10^{10} \, M_\odot$, $\tilde \hbar = 0.02 \, \mathrm{kpc^2 / Myr} = \hbar / (10^{-22} \, \mathrm{eV/c^2})$, $n^p_{tot} \sim 10^{98}$. We simulate the evolution of a system with $M_{tot} = 10^{10} \, M_\odot$, $\tilde \hbar = 0.02 \, \mathrm{kpc^2 / Myr}$, $n^s_{tot} \sim 10^{8}$ and measure the resulting simulated $Q_s(t) \sim 10^{-8} \, t^2$. Using the above relation, this corresponds to a physical $Q_p(t) \sim 10^{-98} t^2$ with corresponding physical breaktime then $t^p_{br} \sim 10^{49} \, \mathrm{Myr}$. Therefore, to relate the simulated $Q_s$ to the physical $Q_p$ we can write

\begin{align} \label{eqn:Q_relation}
    Q_p(t; M_{tot}, \tilde \hbar, n_{tot} = n_{tot}^p) = \frac{n_{tot}^s}{n_{tot}^p} Q_s(t; M_{tot}, \tilde \hbar, n_{tot}^s) \, .
\end{align}

It is important to note that meaningful predictions of the physical breaktime can only be made using $Q_p$. In Figure \ref{fig:Q_n_compare}, we show the physical $Q_p$ predicted by three different simulation sampling $n^s{tot}$. All three simulations give approximately the same prediction for the growth of $Q^p(t)$. We therefore do not expect the prediction of the breaktime to be sensitive to the choice of $n^s_{tot}$ so long as $Q_s(t) \ll 1$.

\begin{figure}[!ht]
	\includegraphics[width = .44\textwidth]{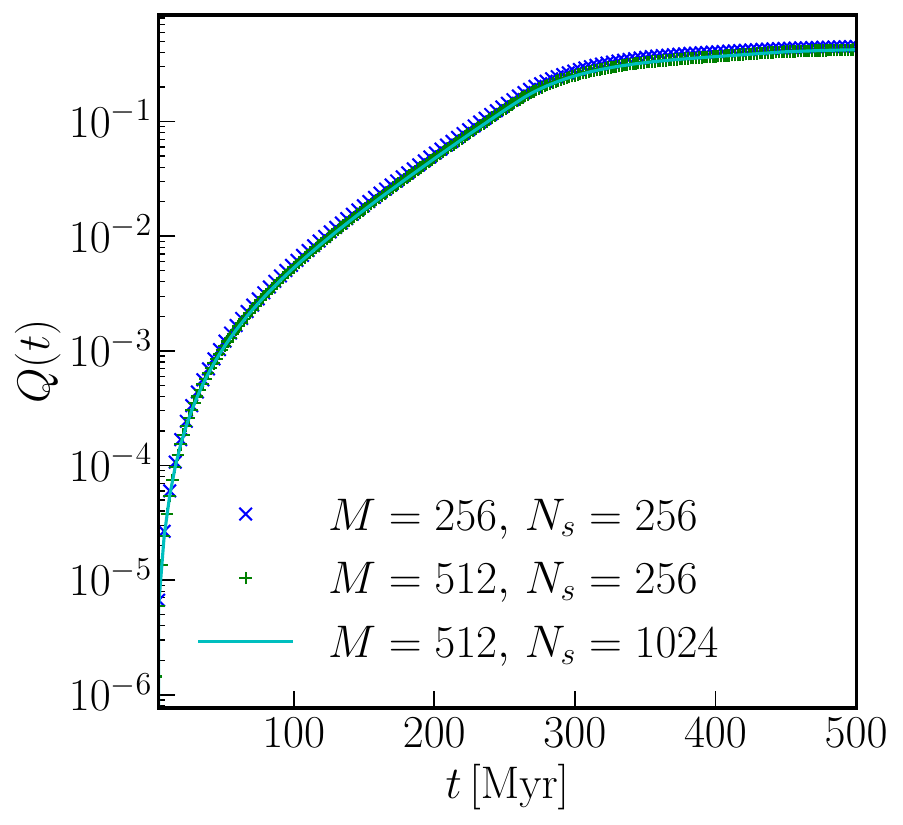}
	\caption{ Here we plot the simulated value of $Q$ for the gravitational collapse of an initial overdensity in a single spatial dimension at three different simulation resolutions. $M$ represents the number of grid cells, and $N_s$ the number of sampling streams. The evolution of $Q$ is the same in all three cases. This demonstrates that the evolution is independent of the specific simulation parameters. For these simulations $M_{tot} = 10^8 \, M_\odot$, $L=60 \, \mathrm{kpc}$, and $\hbar / m = 0.02 \, \mathrm{kpc^2/Myr}$. }
	\label{fig:Q_res_compare}
\end{figure}

Finally, the initial simulated value of $Q^s(t=0)$ will not be exactly $0$ because the sampling scheme provides only an estimation of the underlying distribution. $Q^s(t=0)$ depends on the number of streams $N_s$. We therefore plot simulated $Q$ with this initial value subtracted. The behavior and growth of $Q$ is not sensitive to the choice of $N_s$ so long as $N_s \gg 1$. In Figure \ref{fig:Q_res_compare} we plot the simulated value of $Q$ for simulations with three different resolutions. All three agree, demonstrating again that $Q$ is independent of the resolution parameters $M,\,N_s,$ and $n^s_{tot}$ so long as the resolution is adequate. 

\subsection{Modeling decoherence} \label{subsec:decohere}

Let us start by writing our system as 

\begin{align}
    \ket{A} &= \ket{\rm DM} \ket{\mathcal{E}} \, .
\end{align}

$\ket{\rm DM}$ is the initial quantum state of the dark matter, which we will take to be a coherent state. $\ket{\mathcal{E}}$ is the initial state of the environment, which we will take to be test particles with some well defined positions, $r_i$, and momenta, $p_i$ (though we will later consider that this phase space position can only be known to some resolution). 

We will assume that the gravitational potential is dominated by the dark matter. Using this assumptions we can write

\begin{align} \label{eqn:decohere_hamiltonians}
    &\hat{H}_A = \hat H_{\rm DM} + \hat H_\mathcal{E} + \hat H_{\mathrm{int}} \, ,\\ 
    &\hat H_{\rm DM} = \\ 
    &\int \int dx dy \, \hat \psi^\dagger(x) \frac{-\tilde \hbar^2 \,  \nabla^2}{2} \hat \psi(y) +  \hat \psi^\dagger(x) \, m \, \hat V(x)\, \hat \psi(x) \, ,  \nonumber \\
    &\hat H_{\mathcal{E}} = \sum_i \frac{\hat p_i^2}{2 \, m_p} \, , \\ 
    &\hat H_{\mathrm{int}} = \sum_i m_p \hat V(\hat r_i) \, , \\ 
    &\nabla^2 \hat V(x) = C m |\hat \psi(x)|^2 \, .    
\end{align}
Where $m$ is the mass of the dark matter field, and $m_p$ the mass of the test particles (though this is not actually dynamically relevant as the particle is only coupled through gravity). 

We will write the Wigner function of the dark matter according to the previous sections as an ensembles of streams, and the Wigner function of our test particle as delta functions in momentum-position phase space, i.e. 

\begin{align}
    f_{\rm DM}[\psi, \psi^*] &= \sum_s c_s \delta[\psi - \psi_s(x,t)] \, \delta[\psi^* - \psi^*_s(x,t)] \, , \\
    f_{p}[p, r] &= \sum_{si} \delta[r - r_i^s(t)] \, \delta[p - p^s_i(t)] \, . 
\end{align}
Initially the total Wigner function is just a product $ f_{\rm all} = f_{\rm DM}[\psi, \psi^*] f_{p}[p, r]$, but this will not be the case when entanglement develops, then the state is written $f_{\rm all}[\psi, \psi^*,p, r]$ as a functional of the particle and field configurations.

Using the Hamiltonians in equation \eqref{eqn:decohere_hamiltonians}, the approximate equations of motion for these Wigner functions using the classical equations of motion are

\begin{align}
    \partial_t \psi_s(x,t) =& \frac{-i}{\hbar} \left( \frac{\hbar^2 \, \nabla^2}{2m} + m \, V_s(x) \right) \psi_s \, ,\\
    \nabla^2 V_s(x,t) =& C m \, |\psi_s(x,t)|^2 \, , \\
    \partial_t p^s_i(t) =& -m\, \partial_x V_s(r_i^s,t) \, , \label{eqn:cl_int} \\ 
    \partial_t r^s_i(t) =& \frac{p^s_i(t)}{m} \, .
\end{align}

Equation \eqref{eqn:cl_int}, representing the interaction term in the Hamiltonian, will entangle the particle and field states. The trace over the environment can be evaluated via integral over the test particle phase space positions

\begin{align} 
    f_{DM}^R[\psi, \psi^*] = \int \int \mathcal{D}p \, \mathcal{D}r \, f_{\rm all}[\psi, \psi^*, p, r] \, .
\end{align}

Numerically, the reduced Wigner function can be approximated as
\begin{widetext}
\begin{align}
    &f_{DM}^j[\psi, \psi^*] = \sum_s c_s \, \delta[\psi - \psi_s(x)]\, \delta[\psi^* - \psi^*_s(x)],\, \,\,\, \mathrm{ if } \,\,\, p^s,r^s \in C^j \, , \\
    &f_{DM}^R[\psi, \psi^*] = \sum_j f_{DM}^j[\psi, \psi^*]
\end{align}
\end{widetext}
Where $C^j \subset \mathbb{R}^{2n_p}$ is a region in the $n_p$ test particle phase space configuration space.
$f_{DM}^R[\psi, \psi^*]$ is a classical ensemble of approximately pure state Wigner functions $f_{DM}^j[\psi, \psi^*]$.  Note that we have simplified greatly the overlap between environment configuration states by binning the possible environment phase space configurations and assuming that the overlap is large ($\sim 1$) if two configurations fall in the same bin, $C^j$, and $0$ for configurations in different bins (in general this is only true for well separated bins with appropriate width).  

\section{Test problems} \label{sec:Test_problems_MSM}

In this section we introduce the test problems studied in this paper. The first one is a simple spatial overdensity. The second one we study is the collapse of a random field. Importantly, we study the second test problem in three phases, 1) the initial collapse of the random field into a single virialized object, 2) the stable evolution of this collapsed object, and 3) the merger of two stable collapsed objects. 

\subsection{Sinusoidal overdensity} \label{subsec:sinwave}

We consider the gravitational collapse of an initial overdensity, see figure \ref{fig:1D_density_compare}. In this test problem, an initial perturbation grows exponentially resulting in density shell crossing and a characteristic spiral structure in classical phase space. In corpuscular cold dark matter, this system continues to make smaller scale structures in phase space indefinitely. In the classical field case, the momentum-position uncertainty relation defines a minimum scale under which phase space structure cannot form, see \cite{Eberhardt2020} for a more detailed discussion, resulting in the characteristic ``quantum" pressure associated with this model. We note here that the ``quantum" pressure exists in the purely classical field formalism. 

The initial mean field is given 

\begin{align}
    \psi^{cl}(x_g) &= \sqrt{1 + \delta \, \cos \left( 2 \pi x_g / L \right)} / \mathrm{Norm} \, .
\end{align}

where the norm is chosen such that $\sum_g|\psi(x_g)|^2\, dx = 1$. Recall that when simulating a quantum coherent state, the field instances are chosen normally distributed around the mean field value parameterized by the total number of particles, as

\begin{align} \label{eqn:pickDelta}
    \psi_i(x_g) = \psi^{cl}(x_g) + \delta^R_i(x_g) + \delta^I_i(x_g) i \, , \\
    \delta^R_i(x_g),\, \delta^I_i(x_g) \sim \mathcal{N}(0,\sqrt{1/2}) / \sqrt{n_{tot}} \, .
\end{align}

\subsection{Momentum space gaussian} \label{subsec:boseStar}

\begin{figure}[!ht]
	\includegraphics[width = .48\textwidth]{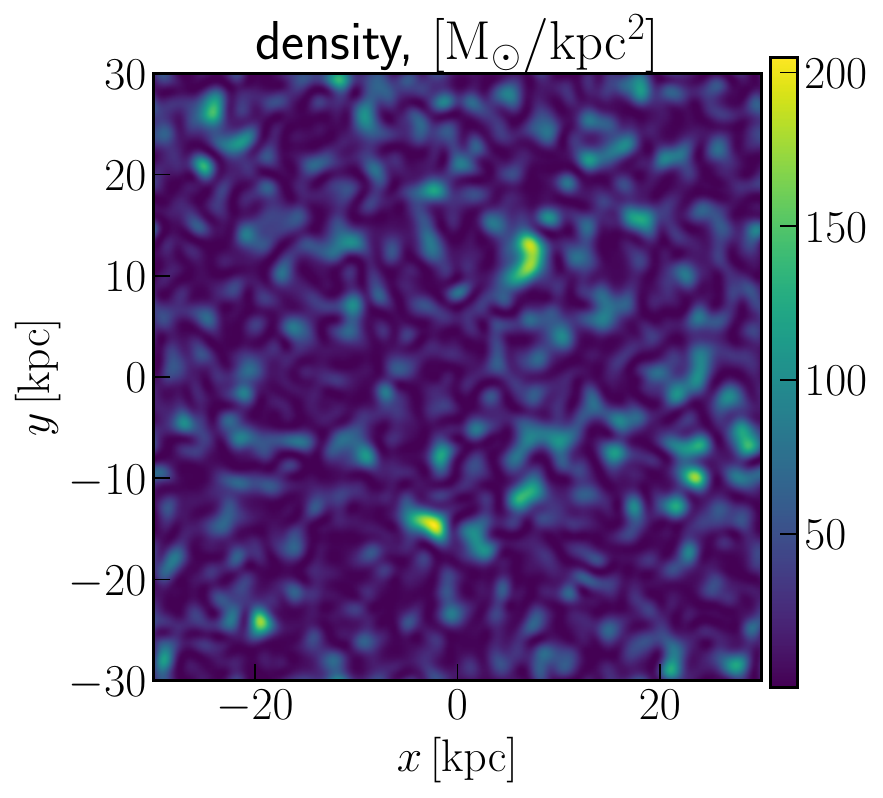}
	\caption{ The density of the initial conditions used to simulate collapsing objects in two spatial dimensions. The momentum density is Gaussian centered on $k = 0$. The granular structure seen in the density is a result of the interference between different momentum modes. Here $M_{tot} = 6 \times 10^{9} \, M_\odot$, $L=60 \, \mathrm{kpc}$, and $\hbar / m = 0.02 \, \mathrm{kpc^2/Myr}$, $t_c \sim 2 \,  \mathrm{Gyr}$, $M = 512^2$.}
	\label{fig:ICs}
\end{figure}

\begin{figure*}[!ht]
	\includegraphics[width = .97\textwidth]{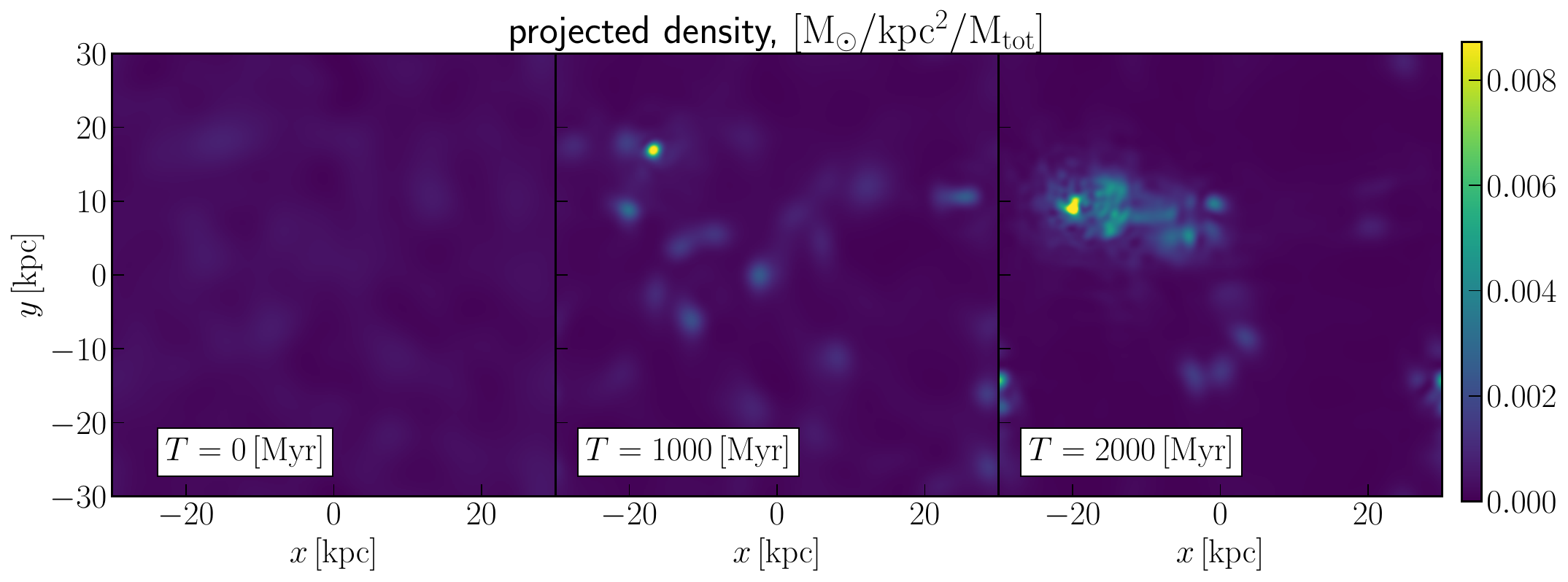}
	\caption{ Here we plot the projected density evolution of the momentum space Gaussian initial conditions described in \ref{subsec:boseStar} in three spatial dimensions. Each column represents a different time, with initial conditions are shown on the left plot and the collapsed object on the right plot. We see the initially randomly distributed granules collapse into an object. Here we set $M_{tot} = 1 \times 10^{10} \, M_\odot$, $\hbar / m = 0.02 \, \mathrm{kpc^2/Myr}$, $L = 60 \, \mathrm{kpc}$, $N = 512^2$, $2 k_d^2 = 0.05 \, \mathrm{kpc^{-2}}$, $t_c \sim 2 \, \mathrm{Gyr}$. }
	\label{fig:3d_collapse}
\end{figure*}
\begin{figure*}[!ht]
	\includegraphics[width = .97\textwidth]{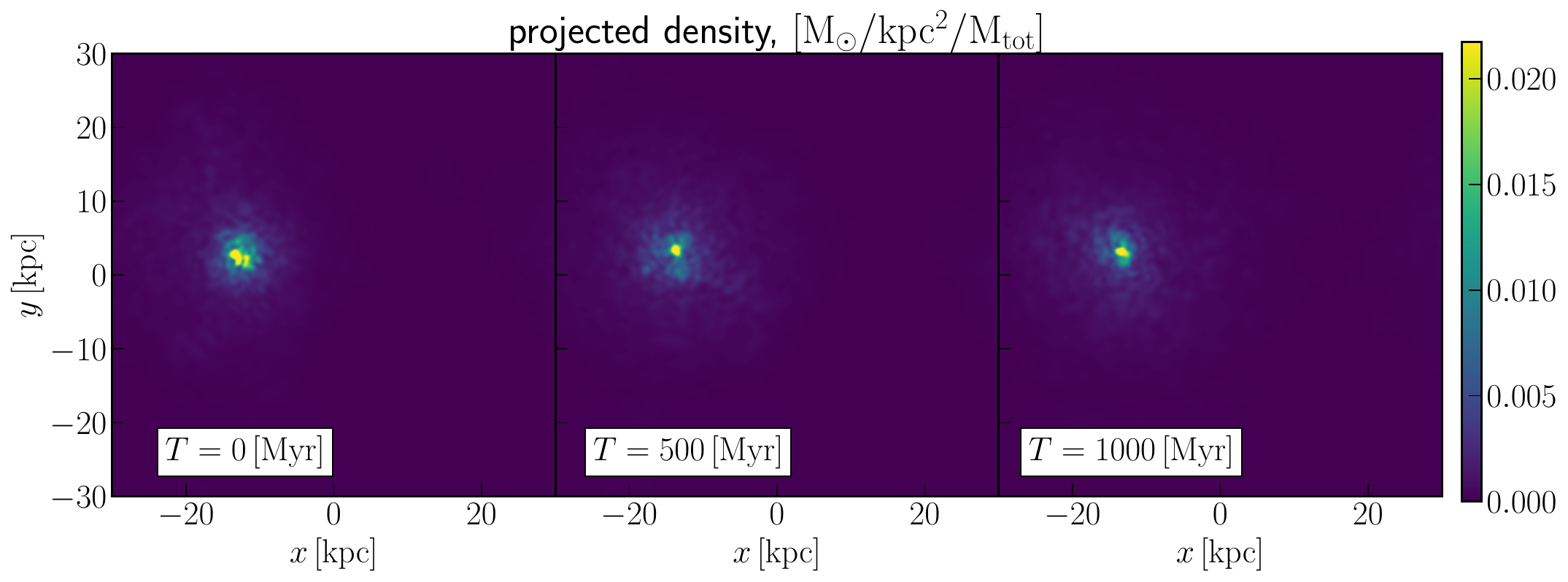}
	\caption{ Here we plot the classical field evolution of the collapsed object resulting from the evolution of the Gaussian momentum distribution described in \ref{subsec:boseStar} in three spatial dimensions. Each column represents a different time, with initial conditions are shown on the left plot and the collapsed object on the right plot. We see the object is supported against further collapse with a continually evolving granular envelope. Here we set $M_{tot} = 10^{10} \, M_\odot$, $\hbar / m = 0.02 \, \mathrm{kpc^2/Myr}$, $L = 60 \, \mathrm{kpc}$, $N = 512^3$. }
	\label{fig:3d_collapsed}
\end{figure*}
\begin{figure*}[!ht]
	\includegraphics[width = .97\textwidth]{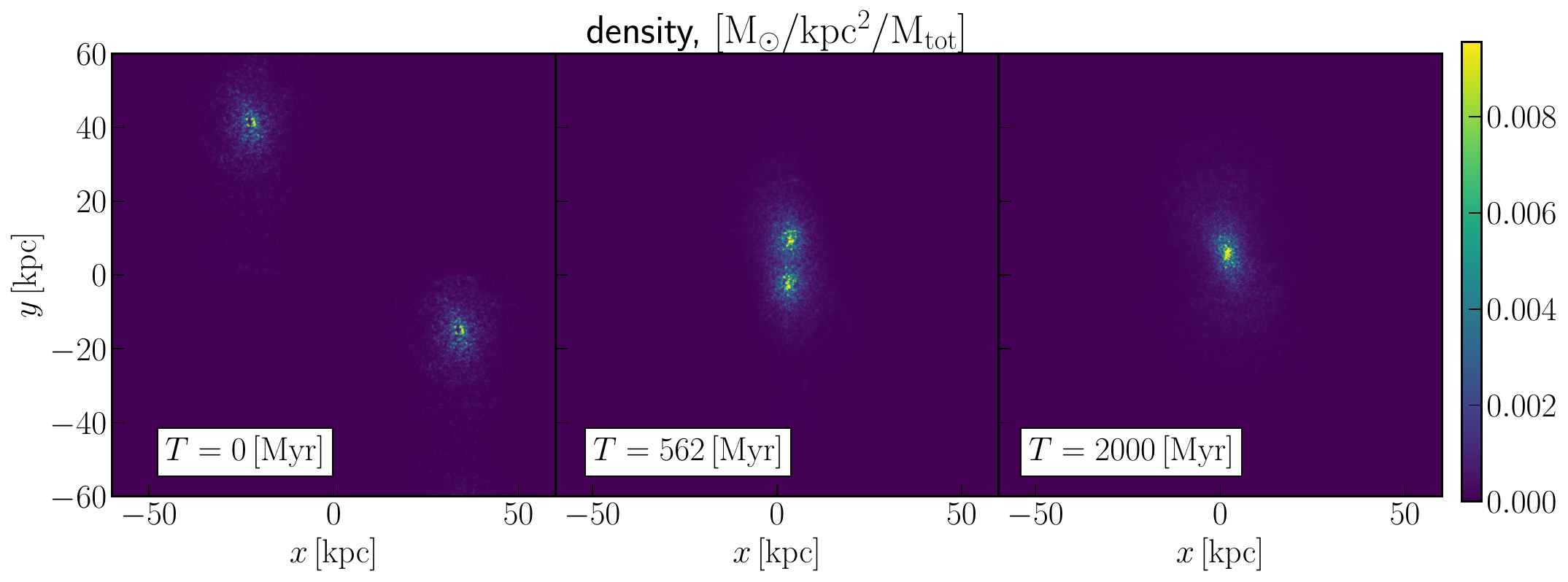}
	\caption{ Here we plot the classical field evolution of the merger of two collapsed object resulting from the evolution of the Gaussian momentum distribution described in \ref{subsec:boseStar} in two spatial dimensions. Each column represents a different time, with initial conditions are shown on the left plot and the collapsed object on the right plot. The collapse time is $t_c \sim 600 \, \mathrm{Myr}$. Here we set $M_{tot} = 1.2 \times 10^{10} \, M_\odot$, $\hbar / m = 0.02 \, \mathrm{kpc^2/Myr}$, $L = 120 \, \mathrm{kpc}$, $N = 1024^2$. }
	\label{fig:2d_merge}
\end{figure*}

The growth and evolution of solitons is one of the most studied systems in ultra light dark matter \cite{Levkov2018, Davies1997, Glennon2022, Tigrak2011,Bar2018,Schive_2016}. It is therefore important to understand how quantum corrections grow in these scenarios. We simulate initial conditions following the example of \cite{Levkov2018} where the initial mean field is chosen in momentum space as

\begin{align} \label{eqn:kspaceGauss}
    \tilde \psi^{cl}(k_i) = e^{-k_i^2/(2 \, k_d^2) + i \tilde \phi_i} \, . 
\end{align}

Where $k_i$ is the wavenumber associated with the $i$th momentum mode. $\tilde \phi_i$ is chosen randomly at each point in momentum space from a uniform distribution, i.e. $\tilde \phi_i \sim U[0,1]$. This describes a Gaussian distributed momentum space with temperature $k_d$. The initial density has granular over-densities given by the interference of momentum modes, see Figure \ref{fig:ICs}. Overtime this system will collapse into a condensed object, see Figure \ref{fig:3d_collapse}, which will then remain a stable ``Bose-star", see Figure \ref{fig:3d_collapsed}. These objects are typically characterized by their granular interference patterns.  

We will be interested in the behavior of quantum corrections during both the collapse phase and the stable object phase. Our sampling scheme assumes that the quantum state is well described by a coherent state at the initial conditions.

We will have three classes of simulations, the first will be starting from the the momentum space Gaussian described in equation \eqref{eqn:kspaceGauss}, i.e. we sample around the classical field shown in Figure \ref{fig:ICs}. These simulations will represent the \textbf{collapsing phase} of the evolution and are intended to demonstrate how quantum corrections grow during gravitational collapse. 

The second class of simulations will start from the collapsed object formed in the classical evolution of the initial conditions described by equation \eqref{eqn:kspaceGauss}, i.e. we sample around the classical field shown in the left panel of Figure \ref{fig:3d_collapsed}, which is the same as the right panel of Figure \ref{fig:3d_collapse}. These simulations will represent the \textbf{post collapsed phase} and are intended to demonstrate how quantum corrections grow in a virialized halo. 

Finally, we will perform simulations taking multiple copies of the collapsed object and allowing them to merge. , i.e. we sample around the classical field shown in the left panel of Figure \ref{fig:2d_merge}, which is four copies of a condensed object given slight perturbations in initial position. These simulations will represent the \textbf{merging of collapsed objects} and are intended to demonstrate how quantum corrections grow during halo mergers.

\section{Results} \label{sec:results_MSM}

In this section, we describe the results of our simulations. We focus this discussion on three main points. First, we estimate the quantum breaktime, i.e. the timescale on which quantum corrections grow large. Second, we estimate the effect of quantum corrections on the dark matter density. Third, we discuss our analysis of the decoherence time for these systems. 

\subsection{Breaktimes}

The breaktime, $t_{br}$ is calculated using the $Q$ parameter, defined in equation \eqref{eqn:Q}. Section \ref{subsec:breaktime} explains the relationship between this parameter and the breaktime in detail. When $Q \sim 1$ the system tends to differ from the predictions of the classical field theory \cite{eberhardt2021Q,eberhardt2021}. The breaktime is estimated by studying how $Q$ grows overtime and estimating when $Q(t_{br}) \sim 1$.

We first note that the result of simulations in higher dimensions generally corroborate the 1D results from \cite{Eberhardt2022}. For the collapse of the Gaussian momentum space density we see an initial quadratic growth of $Q$ followed by an exponential growth during collapse. For already collapsed systems no longer experiencing nonlinear growth we see only a power law growth of $Q$. 

\begin{figure*}[!ht]
	\includegraphics[width = .97\textwidth]{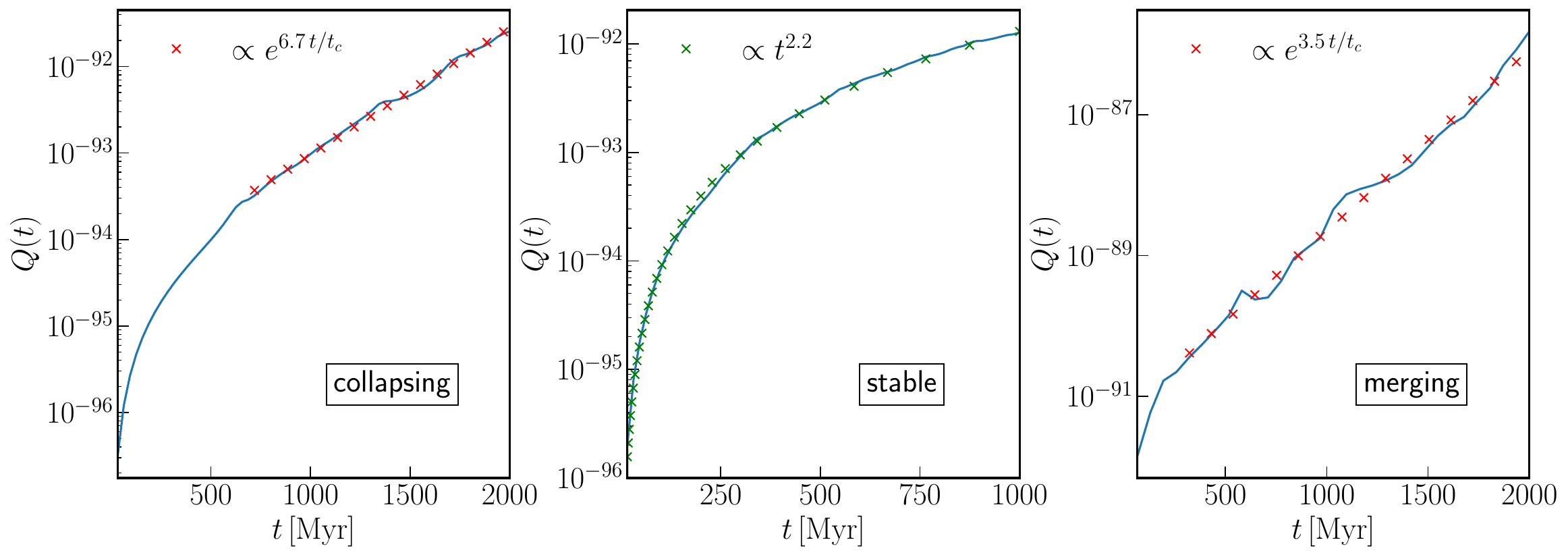}
	\caption{ Here we plot the evolution of the physical $Q$ parameter for the three test problems described in Section \ref{subsec:boseStar}. From left to right the plots show $Q(t)$ for the collapse of a momentum space Gaussian in three spatial dimensions (left), a collapsed object in three spatial dimensions (middle), and the merger of two collapsed objects in two spatial dimensions (right). In both the collapsing and merging case the parameter grows exponentially. In the stable case and in the nonlinear cases at early times the parameter grows quadratically. The growth of $Q$ for these systems corraborates the 1D results found in \cite{Eberhardt2022, Eberhardt_testing}, i.e. that quantum corrections grow exponentially during nonlinear growth and by a powerlaw for virialized systems and at very early times. For the collapsing and stable simulations $M_{tot} = 10^{10} \, M_\odot$, $L=60 \, \mathrm{kpc}$, and $\hbar / m = 0.02 \, \mathrm{kpc^2/Myr}$, $t_c \sim 2$ (approximate collapse time), $\,  \mathrm{Gyr}$, $M = 512^3$. For the merging simulation $M_{tot} = 1.2 \times 10^{10} \, M_\odot$, $L=120 \, \mathrm{kpc}$, and $\hbar / m = 0.02 \, \mathrm{kpc^2/Myr}$, $t_c \sim 0.6 \,  \mathrm{Gyr}$ (approximate merger time), $M = 1024^2$. }
	\label{fig:Q_main}
\end{figure*}
The initial collapse of the k-space Gaussian in three spatial dimensions is shown in Figure \ref{fig:3d_collapse}. The collapse time for this system is approximately $t_c \sim 1 / \sqrt{\rho G} = 1 / \sqrt{G \, 10^{10} M_\odot / 60^3 kpc^3} \sim 2 \,  \mathrm{Gyr}$. We plot the $Q$ parameter for this evolution in the left panel of Figure \ref{fig:Q_main}. 
Like in the 1D results shown in \cite{Eberhardt2022}, the $Q$ parameter initially grows quadratically before growing exponentially. When normalized by the collapse time we see that the exponential growth is comparable to the 1D results \cite{Eberhardt2022}. During collapse 

\begin{align}
    Q(t) \sim \frac{1}{n_{tot}} e^{7 t / t_c} \, .
\end{align}

Therefore the quantum breaktime associated with the nonlinear growth to be 

\begin{align} \label{TbreakEstimate}
    t^{NL}_{br} \sim \frac{\ln(n_{tot})}{7} \, t_c \, .
\end{align}

Note that during the collapse the specific mass of the particle is not relevant as the evolution of the largest scale modes in ULDM is the same as in CDM \cite{Kopp_2017,Eberhardt2020,Hu2000,mocz2019,Schive:2014dra}. 

Figure \ref{fig:2d_merge} shows the evolution of a system in which two collapsed objects are allowed to merge in two spatial dimensions. In this case the quantum corrections grow similarly to the collapsing case, see right panel of Figure \ref{fig:Q_main}, 
i.e. exponentially throughout the merger.

Figure \ref{fig:3d_collapsed} shows the evolution of a system which starts in a collapsed object. This object changes little over the course of the simulation with the main evolution being random changes in the granular structure of the envelope surrounding the solitonic core, behavior typical of collapsed objects in ultra light dark matter. The middle panel of Figure \ref{fig:Q_main}
plots growth of the $Q$ parameter for this system. Unlike the collapsing and merging cases, the growth in well fit by a quadratic growth for teh entire simulation duration. This is explained by the growth of the lowest order terms in the moment expansion around the mean field value, see \cite{Eberhardt2022}. $Q$ is well approximated by 
\begin{align}
    Q(t) = \mathrm{Tr}[\kappa_{ij}] \, t^2 / 2n_{tot} \, ,
\end{align}
where $\kappa_{ij}$ is given from the momentum space mean field values
\begin{align} \label{DefinitionKappa}
     \partial_{tt} \braket{\delta \hat a_i^\dagger \delta \hat a_j} &\sim 2 \mathbb{R}\left[ \sum_{kplbc} \Lambda^{ij}_{pl}  \Lambda^{kj}_{bc} \braket{\hat a_b} \braket{\hat a_c} \braket{\hat a_p^\dagger} \braket{\hat a_l^\dagger} \right]  \nonumber \\
     &\equiv \kappa_{ij} \, .
\end{align}
The corresponding breaktime for the stable system is 
\begin{align}
    t^S_{br} \sim \sqrt{n_{tot}/ \mathrm{Tr}[\kappa_{ij}]} \, .
\end{align}
The difference in the evolution of $Q$ in each of these systems clearly demonstrates that the exponential growth of quantum correction occurs during gravitational collapse and not in already virialized objects. For astrophysical systems $t^{NL}_{br} \ll t^S_{br}$.


\subsection{Large quantum corrections}

\begin{figure*}[!ht]
	\includegraphics[width = .97\textwidth]{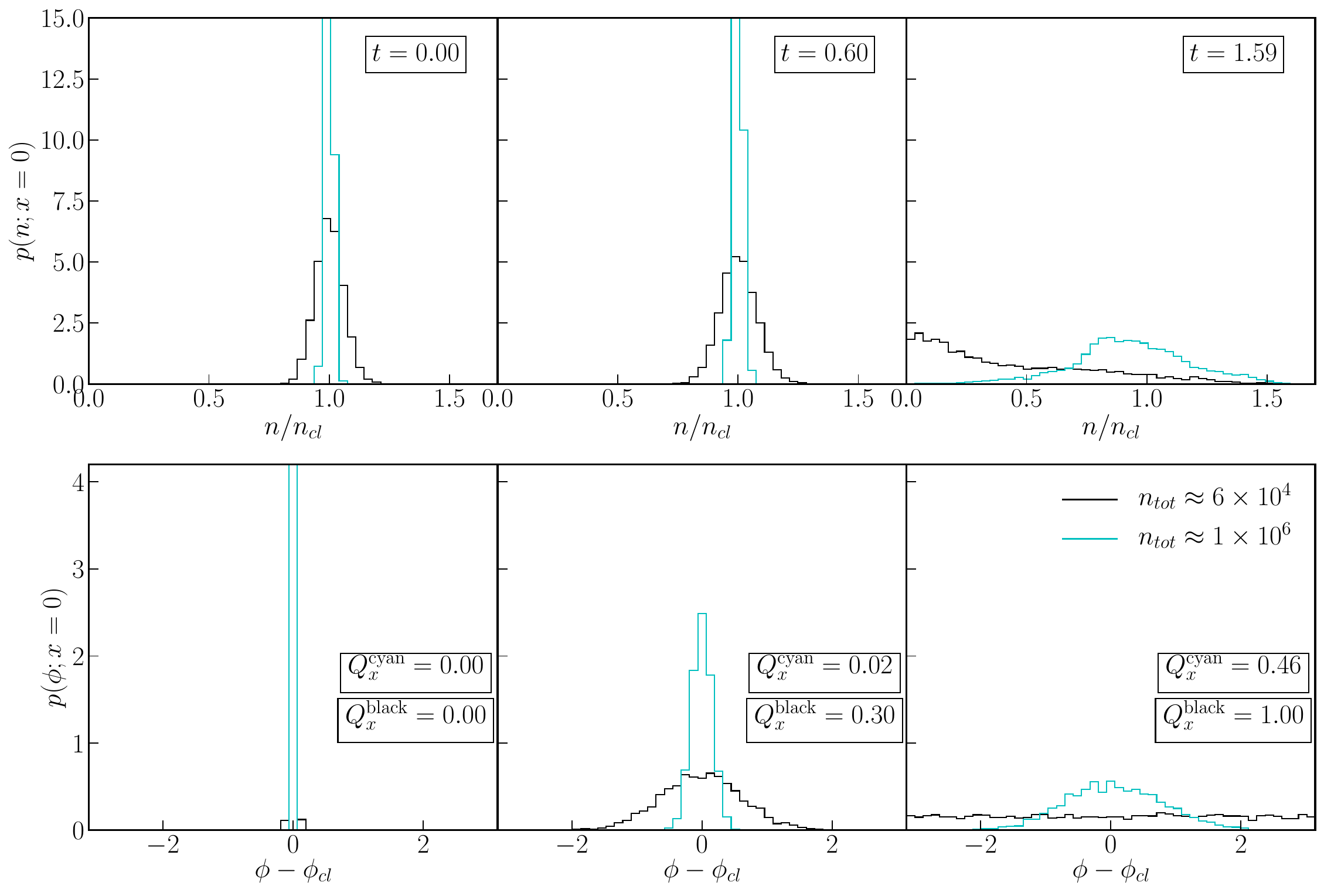}
	\caption{ Here we plot the distribution of complex angles and occupations of the ensemble of fields for a spatial overdensity in one spatial dimension. Two quantum simulations with $n_{tot} \approx 6 \times 10^4$ and $n_{tot} \approx 1 \times 10^6$ are plotted in black and cyan respectively. Each column represents a different time, $t$. The top row shows a histogram of the stream ensemble occupations numbers at $x=0$ and the bottom row a histogram of the stream ensemble complex angles. Shell crossing occurs at $t=1$.  In these simulations $\tilde \hbar = 2.5 \times 10^{-4}$ and $N_s = 1024$. }
	\label{fig:phi_Distr}
\end{figure*}

\begin{figure}[!ht]
	\includegraphics[width = .44\textwidth]{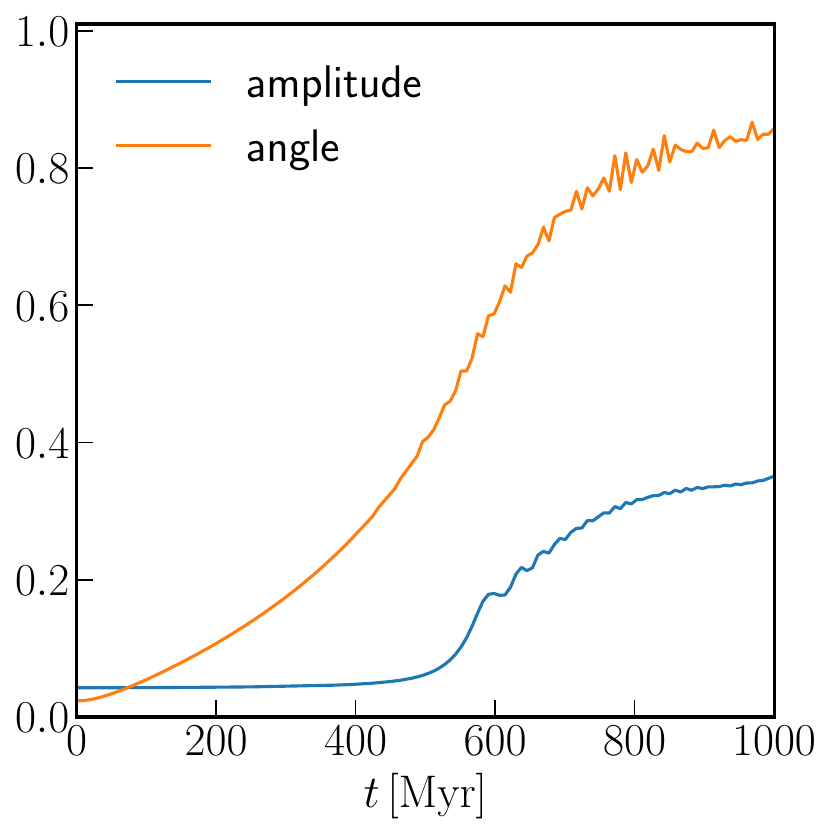}
	\caption{ We plot a normalized mass weighted amplitude variance (see equation \eqref{eqn:varA}) in blue and a normalized mass weighted phase variance (see equation \eqref{eqn:varPhi}) in orange for the gravitational collapse of an over density in a single spatial dimension. We set $N = 512$, $M_{tot} = 10^8 \, \mathrm{M_\odot}$, $L = 60 \, \mathrm{kpc}$, $\hbar/m = 0.01$, $n_{tot} = 10^6$.   }
	\label{fig:var}
\end{figure}

It is instructive to look at the effect of large quantum corrections. This allows us to understand what kinds of predictions we expect to most differ between mean field and quantum systems. In order to demonstrate how these corrections effect the spatial density we have simulated the collapse of a sinusoidal overdensity in one spatial dimension. The results are plotted in Figure \ref{fig:1D_density_compare}. Phase diffusion is the leading order effect, i.e the phase of the wavefunction at a given position becomes less well defined during the collapse. The distribution of occupation numbers and complex angles at $x = 0$ is given for the ensemble of streams in figure \ref{fig:phi_Distr}. In the one spatial dimension case, the corrections grow most quickly during the collapse as opposed to the post collapse virialized stage of the evolution. However, even though the phase is increasingly poorly defined during the collapse, the amplitude of the field is still close to the classical value until shell crossing. If we look separately at the mass weighted fractional amplitude and phase variance, we see that phase variance grows during the collapse, but amplitude variance grows very quickly at shell crossing but slowly before and after, see Figure \ref{fig:var} in which we plot the density weighted amplitude and phase variances, which are given, respectively, as

\begin{align}
    \mathrm{Var}(\tilde A) &= \int dV \, \braket{ \hat \psi^\dagger(x) \hat \psi(x)} \braket{\delta \hat A^2(x)} / \braket{\hat A(x)} \, , \label{eqn:varA} \\ 
    \mathrm{Var}(\tilde \phi) &= \int dV \, \braket{ \hat \psi^\dagger(x) \hat \psi(x)} \braket{\delta \hat \phi^2(x)} / (\pi / \sqrt{3}) \label{eqn:varPhi} \, .
\end{align}

After shell crossing, we can see the primary effect of the quantum corrects is to lessen the degree to which the density has the interference patterns characteristic of scalar field dark matter. This makes sense given that the interference patterns result from differences between well defined spatial phase gradients which become less well defined in the quantum case. 

We expect then that large quantum corrections effect the spatial density by removing the $\sim 1$ fluctuations that come from interference of phase space streams. This results in a reduction of the granular structure typical of collapsed objects in ULDM. 
The result of large quantum corrections on density can be seen for the collapse of a gravitational over-density in a single spatial dimension in Figure \ref{fig:1D_density_compare}, and for the gravitational collapse of an object in two spatial dimensions in Figure \ref{fig:2d_collapseQuantum}. Each shows a reduction in the amplitude of the interference pattern structure of the density. Large quantum corrections therefore affect the density in a way similar to multi-field \cite{Gosenca:2023yjc} or vector-field \cite{Amin2022} ultra light dark matter.

\subsection{Decoherence}

\begin{figure*}[!ht]
	\includegraphics[width = .97\textwidth]{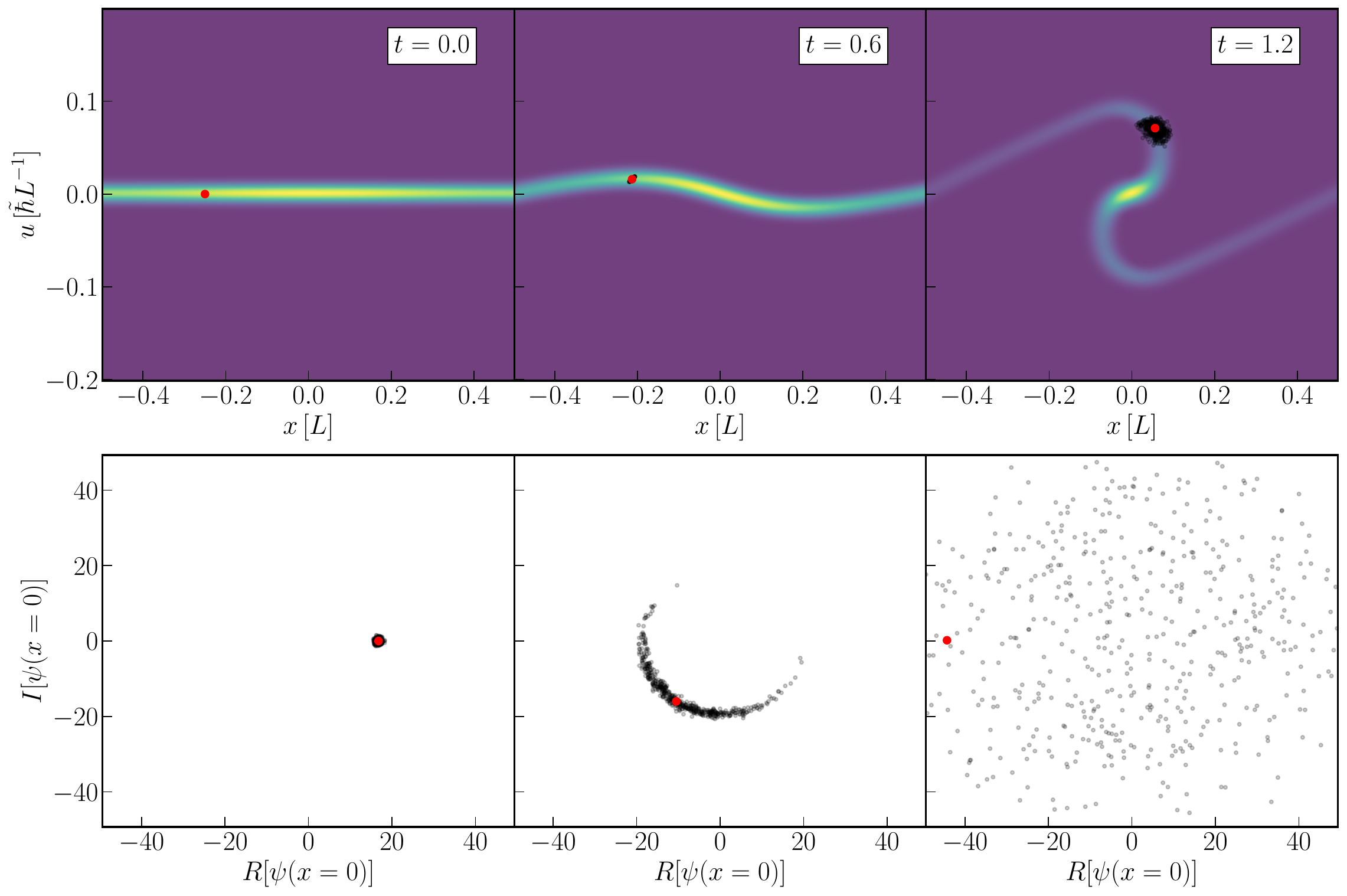}
	\caption{ The evolution of the gravitational collapse of initial overdensity for a quantum coherent state couple dot a tracer particle. The top row shows the phase space of the classical field evolution. Overlayed on top of this is the distribution representing the quantum state of the tracer particle (in black) and the particles classical phase space value (in red). Shell crossing occurs at $t=1$. In the bottom row, we plot the sample streams, used to approximate the Wigner distribution, values at $x=0$ (in black) with the classical field value (in red). We can see that as the quantum state spreads around the classical field value of the field it also spreads around the classical phase space position of the particle. In these simulations $\tilde \hbar = 2.5 \times 10^{-4}$, $n_{tot} \approx 6 \times 10^4$, $M_{tot} = L = 1$, and $N_s = 512$. }
	\label{fig:phaseSpace_decohere}
\end{figure*}

\begin{figure}[!ht]
	\includegraphics[width = .44\textwidth]{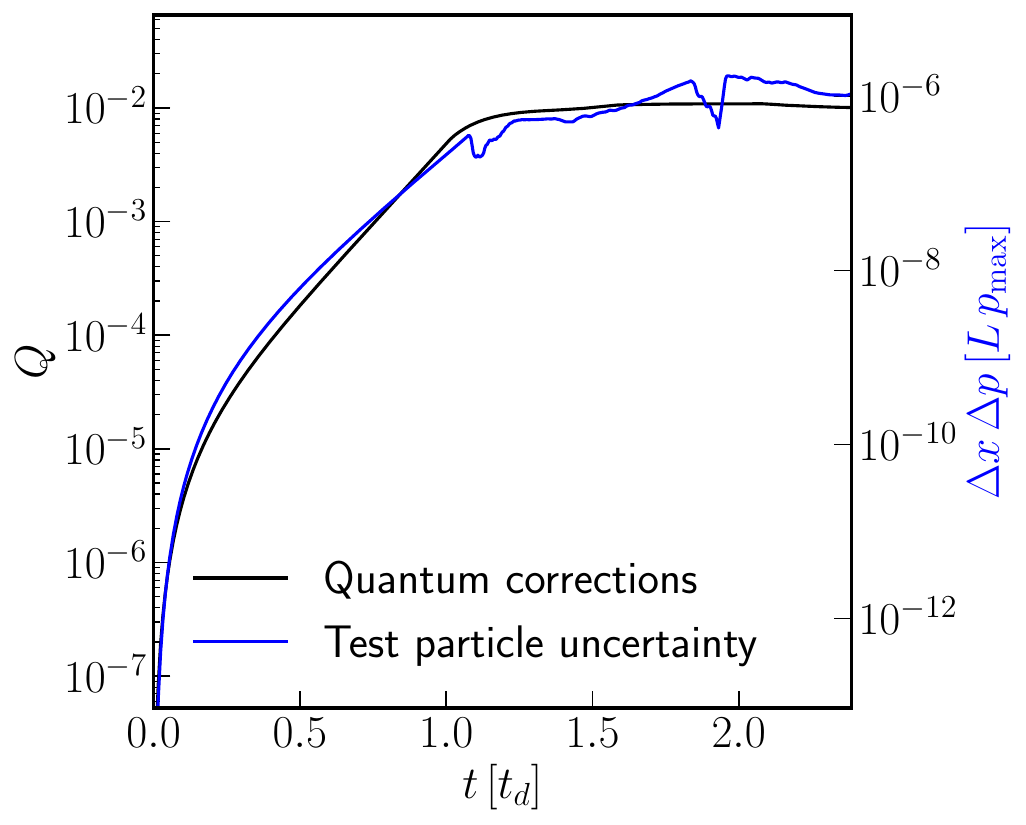}
	\caption{ Here we shows the results of using a test particle to estimate decoherence rates for the collapse of a overdensity in a single spatial dimension coupled to a test particle. The left plot show the uncertainty in the particles phase space position over time, and the right plot shows the $Q$ parameter, measuring the size of quantum corrections in the system. We see that the two grow similarly. Shell crossing occurs at $t=1$.  In these simulations $\tilde \hbar = 2.5 \times 10^{-4}$ and $N_s = 512$, $n_{tot}^s = 6.7 \times 10^7$. }
	\label{fig:decohere}
\end{figure}

Decoherence is tested by coupling a test particle with well defined phase space position to the dark matter Wigner function as described in Section \ref{subsec:decohere}. Baryonic particles take well defined trajectories through phase space, so any quantum effects that quickly puts test particle into macroscopic super positions in phase space is unlike to be stable to decoherence. We simulate the one dimensional collapse of a initial overdensity and place a single test particle at initial position $-L/4$ with initial velocity $0$ and couple it to the dark matter state. We plot the evolution of the system in Figure \ref{fig:phaseSpace_decohere}. In the top row we can see that overtime, the uncertainty in the phase space position of the particle grows as the system collapses. 

We then measure the position-momentum uncertainty of the test particle throughout the evolution, i.e. 

\begin{align}
    \Delta x \, \Delta p = \sqrt{\mathrm{Var}(r^s)\, \mathrm{Var}(p^s)} \, .
\end{align}

The growth of this uncertainty compared to the growth of quantum corrections is plotted in Figure \ref{fig:decohere}. Their growth is very similar. Fundamentally, this is due to the fact that the same dynamics causing the wavefunction to spread around its mean also cause the test particle to spread around its classical phase space value. In order to have large quantum corrections of the kind found here, it is the case that baryonic particles would evolve into macroscopic super positions at the same rate that the quantum corrections develop. Because we observe baryonic particles take well defined, it implies that decoherence rapidly collapsing macroscopic dark matter super positions. 

\section{Discussion} \label{sec:discussion_MSM}

The simulation results allow us to understand a number of quantum effects. We have investigated the timescales on which quantum corrections grow and their effect on observables once they become large. Likewise, we have been able to provide simulations of the decoherence of a test particle. Both of these effects are necessary to understanding the behavior of quantum effects in ultra-light dark matter. 

With respect to the quantum breaktime, the simulations in higher dimensions largely corroborate the 1D results presented in \cite{Eberhardt2022,Eberhardt_testing}. Quantum corrections grow exponentially during nonlinear growth, such as the collapse and merger shown, respectively, in the left and right panel of Figure \ref{fig:Q_main}. 
 The is typical of quantum systems which exhibit classical chaos \cite{Zurek2003}, making this scaling unsurprising for nonlinear gravitational systems. For virialized systems stable against further nonlinear growth, we observe that corrections grow quadratically, see middle panel of Figure \ref{fig:Q_main}. The quadratic growth is seeded from the initial conditions of $Q=0$ by a term proportional to the commutation of the field operators in the field moment expansion of the equations of motion, see \cite{Eberhardt2020, Eberhardt2022}. 

The breaktime for stable systems,
$t^S_{br} \sim \sqrt{n_{tot}/ \mathrm{Tr}[\kappa_{ij}]}$,
is far too long to introduce quantum corrections in the age of the universe for systems with occupations $n_{tot} \sim 10^{100}$. However, the breaktime for nonlinear systems,
\begin{align} \label{TbreakEstimate}
    t^{NL}_{br} \sim \frac{\ln(n_{tot})}{7} \, t_c \, ,
\end{align}
is plausible smaller than the age of the universe of some systems with the shortest dynamical times. This makes an investigation of the effect of quantum corrections and the decoherence timescale useful. 

The predominant effect of quantum corrections is to remove the $\sim \mathcal{O}(1)$ density fluctuations resulting from the coherent interference of streams in phase space, see Figures \ref{fig:1D_density_compare} and \ref{fig:Q_n_compare}. These interference patterns rely on a well defined phase gradient, but the nonlinearity in the Hamiltonian has the effect of causing phase diffusion in the quantum state. When shell crossing occurs and the streams cross each other, the phase is no longer well defined and the amplitude interference pattern is lessened proportionally, see Figure \ref{fig:var}. 

Quantum correction effects on the density fluctuations are similar to multi-field \cite{Gosenca:2023yjc} or vector-field \cite{Amin2022} ultra light dark matter models. Constraints which depend strongly on these fluctuations, such as the heating of ultra-faint dwarf stellar dispersions \cite{Dalal2022,Marsh:2018zyw} and strong gravitational lensing constraints \cite{Powell2023}, would be most impacted when quantum effects are large. Likewise, the effect of large quantum corrections on haloscope experiments was considered in \cite{Marsh:2022gnf}. It should be pointed out that ultra-faint dwarf galaxies has large dynamical times and therefore would have slower grower quantum corrections than large galactic systems. 

An important mitigating factor complicating this story is decoherence. As nonlinearities in the quantum system drive the growth of quantum corrections interactions with the environment project the system into its basis of pointer states. Crucially, the same gravitational interaction which drives the growth of corrections also provides a coupling to the baryonic tracer particle environment. And while it is unlikely that this is the fastest decoherence channel, see for instance \cite{Allali2020,Allali2021,Allali2021b}, any corrections that would force baryonic tracer particles into macroscopic superpositions on phase space cannot be realistic as we can observe that baryonic tracers take well defined paths through phase space. We find that quantum corrections and uncertainty in the tracer's phase space trajectory both grow exponentially, see Figure \ref{fig:decohere}. We can say approximately then that the decoherence timescale is 

\begin{align}
    \tau_d \lesssim t_{br} \, .
\end{align}

This makes sense as both effects are driven by gravitational nonlinearity. Therefore, it is likely that systems which exhibit large quantum corrections also put baryons into macroscopic superpositions which are not observed. This implies that large quantum corrections like the ones we simulate here are unlikely, a result which strongly supports the accuracy of the classical field approximation. Note that this decoherence time does not depend explicitly on the mass like ones previously found \cite{Allali2020,Allali2021,Allali2021b}. Instead it is related to the dynamical timescale, which for this system also describes the timescale on which small perturbations in the initial conditions grow apart in phase space. 

We qualify this support with the following potential caveats. Because the decoherence timescale is fast the pointer states of the system are important to understanding the behavior of the system. If we assume that the pointer states are coherent states then the classical approximation is likely accurate on scales above the scale of quantum fluctuations, i.e. $\sim \mathcal{O}(1/\sqrt{n_{tot}})$. We can say that the pointer states must allow baryonic tracer particles to take well defined trajectories. Naively this means that we would like the state to be an approximate eigenstate of the density operator, but that would be true for coherent states, squeezed states or field number eigenstates, all described by a single classical field but with different quantum properties, or conceivably quantum states that are not described by a single classical field, such as fragmented states (appearing in the context of BECs) where multiple incoherent fields are needed to encapsulate the quantum state. These states have been studied previously in a similar context in for example \cite{kopp2021nonclassicality, Alon2007}. Importantly, a coherent state is only an exact eigenstate of the linear field operator, though the fractional variance of the density operator for a coherent state is on the scale of quantum fluctuations and thus small. It is plausible that other states may be the pointer states. For example, field number states (written in terms of the number eigenstate basis as \footnote{$\ket{\Vec{z}}_f = \sum_{\set{n}} \sqrt{n_{tot}!} \bigotimes_{i=1}^M \frac{ z_i^{n_i}}{\sqrt{n_i!}} \ket{n_i}$}) are eigenstates of the density operator and have been shown in previous work to spread more slowly due to gravitational nonlinearity \cite{eberhardt2021Q}. It is possible that there exist pointer states which satisfy the conditions we describe here but still admit corrections to the classical equations of motion or have interesting quantum properties. 

This work does not contain an analysis of field number states because the Wigner function of the field number state is more difficult to approximate using the truncated Wigner approximation. Likewise, we did not provide any estimation of the pointer states in this work. While Section \ref{subsec:decohere} contains a description of how this method can be used to obtain a reduced density matrix it is unclear to directly identify the pointer states without simply guessing and checking. Finally, we point out that the quartic self-interaction term, which is not considered here, is also unlikely to cause large quantum corrections. Previous work has shown that corrections due to this term grow proportional to a powerlaw \cite{eberhardt2021Q}, similar to the stable systems investigated here. And while this nonlinearity does not drive baryonic tracers into macroscopic superpositions, it is likely to slow to grow quantum corrections in the lifetime of the universe for systems at high occupation. Investigations of field number states, pointer states, and the ultra light dark matter self interactions remain interesting potential future work.

\section{Conclusions} \label{sec:conclusions_MSM}

In this paper we use the truncated Wigner approximation to study quantum corrections to the classical field theory of ultra-light dark matter. We have provided some of the largest and most realistic simulations used to study quantum effects in ultra-light dark matter to date, involving hundreds of modes in 1, 2, and 3 spatial dimensions. Likewise, we have provided the first direct simulations studying quantum decoherence for ultra-light dark matter. Using this approximation we estimate the quantum breaktime for ultra-light dark matter, provide an estimation of the effect of quantum corrections on the density, and investigate decoherence time due to gravitational coupling to a baryonic tracer particle.

Our study of the breaktime corroborates the 1D results in \cite{Eberhardt_testing, Eberhardt2022}. Quantum corrections grow exponentially in systems which grow nonlinearly, and quadratically in stable virialized systems and at very early times. We have now observed these scaling in systems over a wide range of scales, initial conditions, and spatial dimensions, see for example Figure \ref{fig:Q_main}, appendix \ref{app:dimensions}, and systems studied in previous work \cite{eberhardt2021Q,Eberhardt2022,Eberhardt_testing}. 
We find in collapsing systems the breaktime is approximately $t_{br} \sim \frac{\ln(n_{tot})}{7} \, t_c$ where $t_c \sim \sqrt{L^D / G \, M}$ is the dynamic time. 

The systems we have studied in this paper are intended to represent the growth of small scale structure such as dwarf galaxies of approximate mass $M_{tot} \sim 10^{10} M_\odot$ with occupations around $n_{tot} \sim 10^{100}$. Most constraints relating to the impact of ultra light dark matter on structure growth use structure on this scale or smaller. For such a system the exponential growth of quantum corrections we simulated would predict a quantum breaktime $\sim 65 \, \mathrm{Gyr}$. The quadratic growth of quantum corrections results in a much longer breaktime $\sim 10^{45} \, \mathrm{Gyr}$. We note both of these breaktimes are well beyond the age of the universe.

We have found that when quantum corrections are large the leading order effect is to remove the granular structure associated with the $\sim \mathcal{O}(1)$ density fluctuations resulting from interfering streams. This effect can be seen in Figures \ref{fig:1D_density_compare} and \ref{fig:2d_collapseQuantum}. This is similar to the effect of adding additional light fields \cite{Gosenca:2023yjc} or using high spin fields \cite{Amin2022}. Large quantum corrections are therefore most important for studies sensitive to this interference structure such as the heating of dwarf galaxy stellar dispersions \cite{Dalal2022,Marsh:2018zyw}, strong gravitational lensing constraints \cite{Powell2023}, and haloscopes sensitive to the time variation of the field amplitude \cite{Marsh:2022gnf}. 

Our simulation of decoherence indicated that the same perturbations that lead to the spreading dark matter wavefunction would also result in macroscopic phase space super positions of baryonic test particles. Because the same physics governs both processes, this happens at approximately the same rate that quantum corrections grow, see Figure \ref{fig:decohere}. As we do not observe baryonic particles in macroscopic super positions, it is unlikely that a macroscopic super position of dark matter is stable to decoherence. 

These results use direct nonlinear simulations of quantum corrections to provide some of the strongest evidence to date that the classical field approximation used in ULDM simulations is accurate. However, in this work we did not identify the pointer states of the system, study alternative initial quantum states (such as field number states), or take into account ULDM self interaction. These remain interesting potential future work. 

\begin{acknowledgments} 
Some of the computing for this project was performed on the Sherlock cluster. This work was supported by the U.S. Department of Energy under contract number DE-AC02-76SF00515. 
\end{acknowledgments}

\appendix*
\renewcommand{\thesubsection}{\Alph{subsection}}
\section{Appendices}

\begin{widetext}

\subsection{Proof of truncated Wigner method} \label{apndx:proof}

In this appendix we have provide a proof that a classical ensemble of fields, $f_S$,

\begin{align} \label{eqn:ensemble_fS}
    f_S[\psi, \psi^*;t] = \frac{1}{N_s} \sum_i^{N_s} c_i \,  \delta [\psi - \psi_i(x,t)] \, \delta[\psi^* - \psi^*_i(x,t)] \, ,
\end{align}
where

\begin{align}
    \partial_t \psi_i(x,t) &= -\frac{i}{\hbar} \set{H_W[\psi_i(x), \psi_i^*(x)]\, , \, \psi_i(x,t)}_c \, \\
    &= -\frac{i}{\hbar} \frac{\partial H_W[\psi_i(x), \psi_i^*(x)]}{\partial \psi_i^*(x)} \label{eqn:dH_db}
\end{align}
where $H_W$ is the Weyl symbol of the Hamiltonian, solves the equation of motion for this Wigner function, i.e. 

\begin{align} \label{eqn:df_dt}
    \partial_t f_S[\psi, \psi^*;t] \approx -\frac{i}{\hbar}  \, \set{H_W[\psi, \psi^*]\, , \, f_S[\psi, \psi^*;t] }_c.  
\end{align}

We start by taking a time derivative of equation \eqref{eqn:ensemble_fS} and then making substitutions using equations \eqref{eqn:dH_db}, and $a \, \delta(a - b) = b \, \delta(b-a)$ where necessary. We will also use the notational short hand $\delta[\Psi_i] \equiv \delta[\psi(x) - \psi_i(x,t)]$.

\begin{align}
    \partial_t f_S[\psi, \psi^*] &= \frac{1}{N_s} \partial_t \sum_i c_i\, \delta [\Psi_i] \, \delta[\Psi_i^*] \nonumber \\ 
    &= \frac{1}{N_s} \sum_i c_i \left( \partial_t \delta [\Psi_i] \right)  \, \delta[\Psi_i^*] + \delta [\Psi_i] \, \left( \partial_t \delta[\Psi_i^*] \right) \nonumber \\
    &= \frac{1}{N_s} \sum_i c_i  \left(  \frac{\partial \delta [\Psi_i]}{\partial \psi} \frac{\partial \psi_i(x,t)}{\partial t} \right)  \, \delta[\Psi_i^*] + c.c. \nonumber \\
    &= \frac{1}{N_s} \sum_i c_i \left(  \frac{\partial \delta [\Psi_i]}{\partial \psi} \frac{\partial \psi(x,t)}{\partial t} \right)  \, \delta[\Psi_i^*] + c.c. \nonumber \\
    &= -\frac{i}{\hbar} \frac{1}{N_s} \sum_i c_i \left(  \frac{\partial \delta [\Psi_i]}{\partial \psi} \frac{\partial H_W[\psi(x), \psi^*(x)]}{\partial \psi^*} \right)  \, \delta[\Psi_i^*] - c.c. \nonumber \\
    &= -\frac{i}{\hbar} \frac{\partial H_W[\psi(x), \psi^*(x)]}{\partial \psi^*}   \frac{\partial }{\partial \psi} \frac{1}{N_s} \sum_i c_i \, \delta [\Psi_i]  \, \delta[\Psi_i^*] - c.c. \nonumber \\
    &= -\frac{i}{\hbar} \frac{\partial H_W[\psi(x), \psi^*(x)]}{\partial \psi^*}   \frac{\partial f_S[\psi(x), \psi^*(x)]}{\partial \psi}  - c.c. \nonumber \\
    &= -\frac{i}{\hbar}  \, \set{H_W[\psi(x), \psi^*(x)]\, , \, f_S[\psi(x), \psi^*(x)] }_c. \label{eqn:proof_end}
\end{align}

And we see that equation \eqref{eqn:proof_end} is the same as equation \eqref{eqn:df_dt}, completing the proof. 

\end{widetext}

\subsection{MSM: A Rust/C++ Implementation} \label{appendixB}
We use \textsc{Rust} bindings for the \textsc{C++} library \textsc{Arrayfire} to create a fast single-GPU implementation of the ensemble method presented in this paper. We call this implementation MSM: Multi-Stream Method. The implementation can be found at \hyperlink{https://github.com/andillio/MSM}{https://github.com/andillio/MSM}.

The implementation includes several test problems such as the spherical tophat, coherent and incoherent Gaussians, and supports user-specified initial conditions. 
Since \textsc{Python} remains a popular language, the code outputs snapshots in \textsc{Numpy}'s npy format for ease of use and allows for such format to be read in as initial conditions. The implementation supports one, two, and three spatial dimensions. The repository also includes a synthesizer tool, which synthesizes the streams output by the simulator. It executes and averages arbitrary functions $\mathbb{C}^{N^3} \to \mathbb{C}^{N}$ and $\mathbb{C}^{N^3} \to \mathbb{C}$ across the streams, where $N$ is the number of spatial cells in the individual streams 
These functions can be applied either on the individual streams before being averaged or on the averaged wavefunction. Averaging the stream wavefunctions, their Fourier transforms, and their respective squares, along with calculating the $Q$ used in this paper are several examples of this use.

\begin{figure}[!ht]
	\includegraphics[width = .44\textwidth]{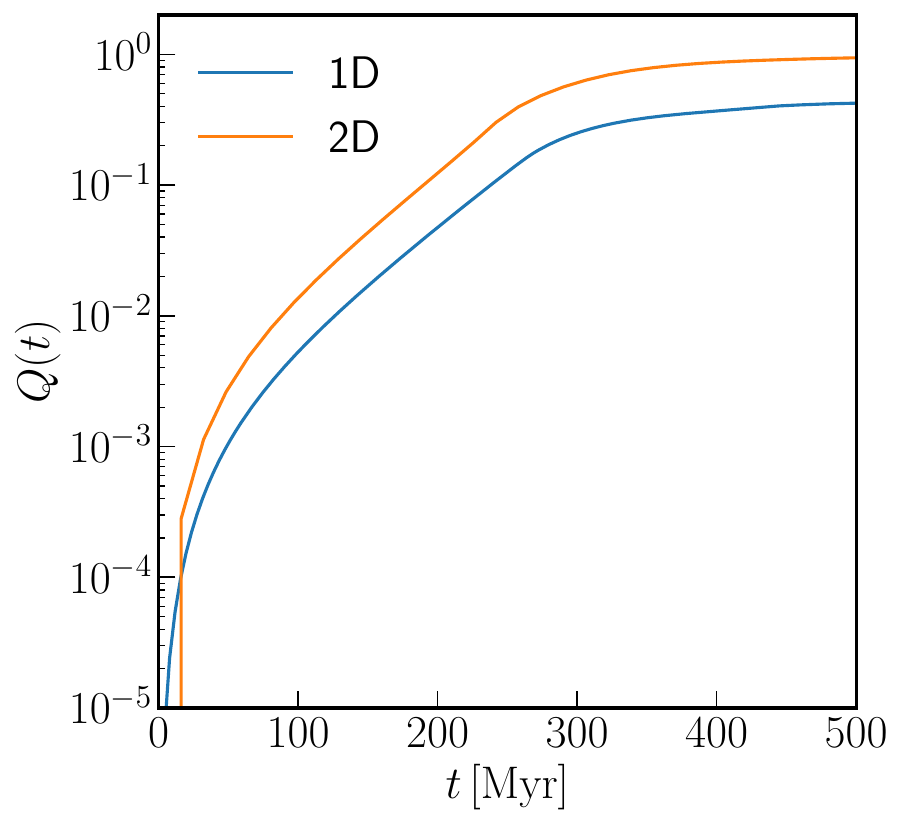}
	\caption{ Here we plot $Q(t)$ for the gravitational collapse of an initial over-density in a single and two spatial dimensions for two systems with the same dynamical times. The evolution is quite similar in both cases, reproducing the familiar results seen for this test problem \cite{Eberhardt2022}. For the 1D sim we set $N = 512$, $M_{tot} = 10^8 \, \mathrm{M_\odot}$, $L = 60 \, \mathrm{kpc}$, $\hbar/m = 0.01$, $n_{tot} = 10^6$. For the 2D sim we set $N = 512^2$, $M_{tot} = 10^8 * L \, \mathrm{M_\odot}$, $L = 60 \, \mathrm{kpc}$, $\hbar/m = 0.01$, $n_{tot} = 10^6$.   }
	\label{fig:Qsine}
\end{figure}

\subsection{Comparing spatial dimensions} \label{app:dimensions}

We include a breif study of systems with different numbers of spatial dimensions to demonstrate that the behavior we observe here is not specific to any particular set of dimensions. Simulations of the same test problem in higher dimensions produce similar results. For example, in Figure \ref{fig:Qsine}, we show the evolution of the $Q(t)$ parameter for the collapse of a sinewave overdensity in a single compared with two spatial dimensions with the same total occupation. We can see that the evolution is quite similar qualitatively, the only difference being the a factor of about $2.5$ in the value of $Q$, a factor which has a vanishingly small effect on the order of the quantum breaktime. This corroborates the results of this work, in which the results found in \cite{Eberhardt2022} in a single spatial dimension are largely applicable in higher dimensions and with a different numerical method. 

\bibliography{BIB}

\end{document}